\documentclass[aps,prx,twocolumn,reprint,longbibliography]{revtex4-2}
\usepackage{graphicx,color}
\usepackage[utf8]{inputenc}
\usepackage[T1]{fontenc}
\usepackage[colorlinks=true,linkcolor=blue,citecolor=blue,urlcolor=blue]{hyperref}
\usepackage{amsmath}
\usepackage{amsthm}
\usepackage{amssymb}
\usepackage{physics}
\usepackage{amsfonts}
\usepackage{comment}
\usepackage[normalem]{ulem}
\newcommand{\be}{\begin{equation}} 
\newcommand{\ee}{\end{equation}} 
\newcommand{\bea}{\begin{eqnarray}} 
\newcommand{\eea}{\end{eqnarray}}

\begin{document}
\title{Statistical Uncertainties  of Limit Cycle Systems in Langevin Bath}
\author{Dipesh K. Singh} \email{dks20ms176@iiserkol.ac.in} \author{P. K. Mohanty}
\email{pkmohanty@iiserkol.ac.in}
\affiliation {Department of Physical Sciences, Indian Institute of Science Education and Research Kolkata, Mohanpur, 741246, India.}
\date{\today}

\begin{abstract}
We show that limit cycle systems  in Langevin bath    exhibit uncertainty  in  observables  that define the  limit-cycle plane, and  maintain a positive lower bound.   The uncertainty-bound depends on the parameters  that  determine the  shape and periodicity of the limit cycle.  In one dimension, we  use the framework of canonical dissipative systems  to  construct  the limit cycle,  whereas  in two dimensions, particle in central potentials   with  radial dissipation provide us  natural examples. We also   investigate   how uncertainties, which are absent in deterministic systems, increase  with time  when the systems are attached to a bath  and eventually cross   the  lower bound  before reaching  the steady state.
\end{abstract}
\maketitle

%\begin{quotation}
%We explore how systems that sustain limit cycle oscillations are  influenced by thermal noise, in a Langevin bath. These systems undergo   noisy motion around  the limit cycle  generating  fluctuations  in  observables  defining their plane of motion. It turns out that these fluctuations  maintain a non-trivial  lower bound  with their  product  depending only on factors that shape the cycle  and determine its periodicity. In  one-dimension,  the  limit cycles are  created   using  the method of canonical dissipation. In   two dimension they appear  naturally for particles  in a  radially damped  central force field with  limit cycles produced in position-position, position-momentum and  momentum-momentum phase planes; interestingly   the uncertainty   in   position-momentum  finds   half the  magnitude of  angular momentum as its lower bound.

%We also study, how uncertainties are generated from any specific   initial state   with  zero  uncertainty and eventually cross  the minimum bound before the system settles into its steady state.
%\end{quotation}

\section{\label{intro:level1}Introduction}
A limit cycle is a closed trajectory in the phase space of a dynamical system \cite{Christopher2007, Strogatz2018}, and its stability depends on whether neighboring trajectories approach the limit cycle as time progresses. Limit cycle (LC) systems are abundant in nature and play a fundamental role in  predicting   and studying  self-sustained oscillations   observed  across various scientific disciplines, such as neuroscience \cite{Delcomyn1980, Pals2024}, ecology \cite{Gilpin1975,Abrams1998}, electrical engineering \cite{Witthaut2022},  and  chemical systems \cite{Erban2023, Field1974}. Ranging from the Hodgkin-Huxley model \cite{Hodgkin1952} that demonstrates how neurons exhibit rhythmic firing patterns,  periodic population fluctuations in predator-prey dynamics \cite{May1972} described by the Lotka-Volterra equations, application of van der Pol oscillators \cite{vanderPol1927} in radio frequency applications, aircraft wing-fluttering \cite{Carroll1982} and nonlinear oscillations in electrical circuits \cite{Witthaut2022}, limit cycle oscillations play a key role.  The emergence of stable limit cycles is an essential phenomenon that sustains life on earth \cite{Roenneberg2008}: beating of the heart, the circadian rhythms, regulation of biological pathways, planetary and climate cycles are oscillatory processes, and their emergence and stability are essential in the continuity of life-processes. 

Stability of LCs   to different  kind of perturbations  and  noise  has been   a question of  importance  for decades \cite{Kurrer1991, Shmakov2024, Sarkar2024}.  Nonlinearity, which is 
responsible for giving birth to  LCs  can  also destroy them \cite{Strogatz2018}.  The  effect of  noise is usually destructive; small amount of noise can  produce irregular LCs with trajectories not so far  away from the original LC curve, whereas strong noise can destroy the LC behaviour \cite{Kurrer1991, Shmakov2024, Sarkar2024}. Coupled limit cycle systems  may exhibit   entrainment  in presence of noise \cite{Mitarai2013}. A recent study  shows that   moderate noise on a limit cycle oscillator  can produce  counterrotation and bistability  \cite{Newby2014}.
%observed in dissipatively coupled condensate\cite{Kim2020}, semiconductor Optical Cavities \cite{Hamerly2015}.\\

In this article, we show that Limit cycle  systems in Langevin bath exhibit unusual residual fluctuations that survive in the zero-noise limit. As a consequence position-momenta uncertainties, i.e., a positive lower bound in $\Delta x \Delta p_x,$ is observed when limit cycles are formed in  $x$-$p_x$ phase plane. The lower-bound depends on the parameters that control the size and shape of the limit cycle and the time required for the system to complete one cycle. For a large set of central potentials in two dimensions, with conserved angular momentum  $\ell$, we find that $\Delta x \Delta p_x \ge \frac{|\ell|}{2}$.

%{\clr change- }In  Fig. \ref{fig:distribution_sho} we demonstrate these results for a  simple harmonic oscillator (SHO) in  two dimension (2D) in presence of a  radial Langevin  bath at temperature $T,$  that keeps the angular momentum $\ell$  conserved. The steady state marginal distribution $P(x,y)$, shown in   Fig. \ref{fig:distribution_sho}, is  centered around  the curve  $x^2+y^2=r_c^2;$ the noisy circular motion   about  this curve in real space  translates to similar motion in $x$-$p_x$ phase plane  resulting   in  a position momenta uncertainty  relation $\Delta x \Delta p_x \ge |\ell|/2.$

\begin{figure}[t]
    \centering
    \includegraphics[width=\linewidth]{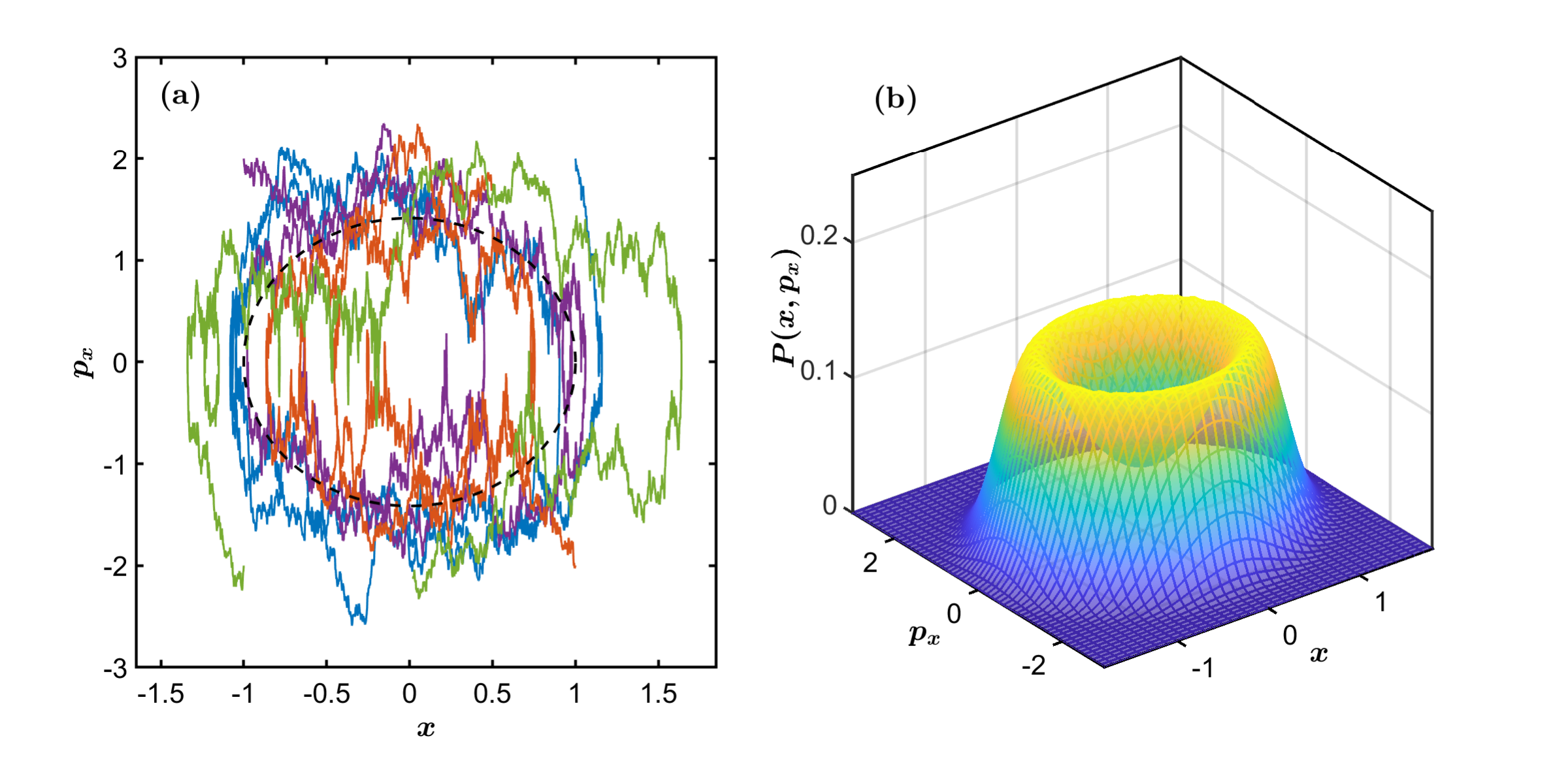}
    \caption{(a) Four ample paths (solid lines) of a  limit cycle  system constructed from a 1D simple  harmonic oscillator  following Eq.  \eqref{eq:sho1d}  with $k=2=\gamma, E_0=1,T=0.5,$   along with the limit cycle (dashed lines). (b) Steady state distribution in the $x$-$p_x$ plane.}
    \label{fig:distribution_sho}
\end{figure}

 In  Fig. \ref{fig:distribution_sho}(a) we demonstrate these results for a limit cycle constructed out of a simple harmonic oscillator (SHO) in  one dimension (1D) using the framework of canonical dissipation. The steady state distribution $P(x,p_x)$, shown in   Fig. \ref{fig:distribution_sho}(b), is  centered around  the limit cycle trajectory $p_x^2/2+\omega_0^2x^2/2=E_0$; the noisy elliptical motion   about  this curve results in  a position momenta uncertainty  relation $\Delta x \Delta p_x \ge E_0/\omega_0.$

Thermodynamic uncertainty relations \cite{Dieball2023, Kwon2022} or statistical bounds \cite{Sabbagh2024}  are not new  to statistical physics, but their presence in  limit cycle systems is rather surprising. It is interesting  to  note  that  some of the  features of a two dimensional harmonic oscillator  in the presence of a moderate thermal noise  is  quite similar  to what one observes in the corresponding quantum system. In particular,  we find that the  position-momentum   uncertainty    bound is   proportional to the conserved angular momentum  of the system and the position autocorrelation at  low temperature  exhibits sinusoidal oscillations.

The article is organized as follows: we start with one dimensional systems in Sec. \ref{1d_LC:level1}, highlighting a way of generating limit cycles by modifying Hamiltonian dynamics in Sec. \ref{1d_LC_construction:level2} and introduce the notion of uncertainty using an example in Sec. \ref{1d_LC_sho:level2}. We move on to two dimensional systems in Sec. \ref{2d_LC:level1} where we present a method of generating limit cycles using central potentials (Sec. \ref{2d_LC_centrlpot:level2}) which exhibit uncertainty relations. We then elucidate this general framework by considering some examples in Secs. \ref{2d_LC_r_alpha:level2} and \ref{2d_LC_log}. We finally highlight the dynamical features of these systems via numerical simulation in Sec. \ref{sec:dynamical} followed by a rich discussion in Sec. \ref{sec:discussion}. The conclusion with a few remarks are given in Sec. \ref{sec:conclusion}.

\section{\label{1d_LC:level1}Limit Cycles in One Dimensional Systems}
Time independent Hamiltonian systems in one dimension (1D) with added linear damping cannot produce limit cycles in the $x$-$p_x$ phase plane. In this section, we utilize the framework of canonical dissipative systems \cite{Haken1973,Schweitzer2001,Ebeling2005} to modify Hamiltonian dynamics which yields limit cycles, and show the presence of position-momentum uncertainties.
 
\subsection{\label{1d_LC_construction:level2} Constructing Limit Cycles}
Any Hamiltonian, $H(x,p_x)$, which produces closed orbits given by $H=E_0$ in the $x$-$p_x$  phase space can be used to construct limit cycles in the same phase plane under the framework of canonical dissipative systems. We then use a Langevin-like bath. The noisy dynamics of the limit cycle oscillator is described by 
\begin{equation}
    \dot{x}=\pdv{H}{p_x} \qcomma \dot{p}_x=-\pdv{H}{x}-\gamma g(H)\pdv{H}{p_x}+\sqrt{\gamma T}\xi(t) \label{eq:Lang_1d}
\end{equation}
where $\gamma>0$ and $\xi(t)$ is a Gaussian white noise with $\ev{\xi(t)}=0$ and $\ev{\xi(t)\xi(t')}=2\delta(t-t')$. We use units where mass and Boltzmann constant are unity $(m=k_B=1)$. A non-decreasing function $g(H)$, with appropriate dimension, is chosen such that $g(H)=0$ describes a closed curve in the phase space. Then, in absence of noise (at $T=0$) the above system sustains limit cycle oscillations, whose equation in the phase space is given by $g(H)=0$. The Fokker-Planck equation can be written as \cite{supp}
\begin{equation}
    \pdv{P}{t}=[H,P]+\pdv{p_x}(\gamma g(H)\pdv{H}{p_x}P+\gamma T\pdv{P}{p_x})
\end{equation}
where $[H,P]=\pdv{H}{x}\pdv{P}{p_x}-\pdv{H}{p_x}\pdv{P}{x}$ is the Poisson bracket of $H$ and $P$. The above equation is easily solved using the ansatz $P=P(H)$ leading to a steady state distribution
\begin{equation}
    P(x,p_x)=\dfrac{1}{\mathcal{Z}}\exp(-\beta\int\dd{H(x,p_x)}g(H(x,p_x))) \label{eq:1d_steadystate}
\end{equation}
where $\beta\equiv T^{-1}$ and $\mathcal{Z}$ is the partition function,
\begin{equation}
    \mathcal{Z}(\beta)=\int_{-\infty}^{\infty}\dd{p_x}\int_{-\infty}^{\infty}\dd{x}\exp(-\beta\int\dd{H}g(H)).  \label{eq:partitionfunction_1D}
\end{equation}
The validity of the distribution requires the probability density integrated over all space, that is, the partition function $\mathcal{Z}$, to be finite.

The simplest $g(H)$ possible is a function linear in $H$, given by $g(H)=H/E_0-1$, $E_0>0$. In the absence of noise (at $T=0$), this system produces a limit cycle whose equation in the phase space is given by $H=E_0$. The  solution of the Fokker-Planck equation leads to a steady state distribution
\begin{equation}
 P(x,p_x)=\frac{1} {\mathcal{Z}_{E_0}} \exp(-\dfrac{\beta}{2E_0}(H(x,p_x)-E_0)^2), \label{eq:linear_g_ss}
\end{equation}
which is a valid probability density function (PDF) if $\mathcal{Z}_{E_0}$ is finite.

Here, the inhomogeneous dissipation creates a nonequilibrium bath leading to a steady state that differs from  Boltzmann distribution with respect to $H;$ $T$ being only a parameter of the bath, need not be treated as temperature. Such a nonequilibrium bath is a necessity in one dimension, where linear damping cannot generate a limit cycle.

\subsection{\label{1d_LC_sho:level2}Limit Cycle construction using 1D Simple Harmonic Oscillator}
We consider one of the simplest Hamiltonian systems, a one dimensional simple harmonic oscillator, described by the Hamiltonian, $H=p_x^2/2+kx^2/2$, with frequency $\omega_0=\sqrt{k}$. It produces elliptical orbits in the $x$-$p_x$ phase space. We also choose $g(H)=H/E_0-1$, $E_0>0$. The noisy dynamics of the limit cycle oscillator is described according to \eqref{eq:Lang_1d},
\begin{equation}
    \dot{x}=p_x \qcomma \dot{p}_x=-kx-\gamma\qty(\dfrac{H}{E_0}-1)p_x+\sqrt{\gamma T}\xi(t) .\label{eq:sho1d}
\end{equation}
As expected, in the absence of noise (at $T=0$) the above system produces an elliptical limit cycle whose equation in the phase space is given by $H=E_0$. The solution of the Fokker-Planck equation corresponding to the Langevin dynamics in Eq. \eqref{eq:sho1d} leads to a steady state distribution given by Eq. \eqref{eq:linear_g_ss},
\begin{equation}
 P(x,p_x)=\frac{1} {\mathcal{Z}_{E_0}} \exp(-\dfrac{\beta}{2E_0}\qty(\dfrac{p_x^2+kx^2}{2}-E_0)^2), \label{eq:steady_1dsho}
\end{equation}
with the partition function \cite{supp}, $\mathcal{Z}_{E_0}$,
\begin{equation}
    \mathcal{Z}_{E_0}(\beta)=\sqrt{\dfrac{2E_0}{k\beta}} \pi^{3/2}\qty(1+\mathrm{erf}\qty(\sqrt{\beta E_0/2})).
\end{equation}

Let's  proceed to calculate  the moments. As  expected $\ev{x}=0=\ev{p_x},$ and 
\begin{align}
    k\ev{x^2}=\ev{p_x^2}=\ev{H}=E_0+\dfrac{e^{-\beta E_0/2}\sqrt{2E_0/\pi\beta}}{1+\mathrm{erf}\qty(\sqrt{\beta E_0/2})}.
\end{align}
Now we turn our attention to the uncertainty, which is defined for an observable $O$ as $\Delta O\equiv\sqrt{\ev{O^2}-\ev{O}^2};$ in steady state we get,
\begin{equation}
    \Delta x\Delta{p_x}=\dfrac{1}{\omega_0}\qty(E_0+\dfrac{e^{-\beta E_0/2}\sqrt{2E_0/\pi\beta}}{1+\mathrm{erf}\qty(\sqrt{\beta E_0/2})})
\end{equation}
which does not vanish in  the  $T\to 0$ limit, leading to a position-momentum uncertainty relation,
\begin{equation}
    \Delta x\Delta{p_x}\geq\dfrac{E_0}{\omega_0}. \label{eq:1d_uncertainty}
\end{equation}
The equality holds only in the limiting sense, that is, as $T\to0$. Note that $\ev{H}(T\to0)=E_0$, which is the equation of the limit cycle in the $x$-$p_x$ plane.

We claim that such uncertainty relations are common to limit cycle systems in Langevin bath. We shall always have some uncertainty relation whenever there is a limit cycle in any two dimensional projection of phase space. To study higher dimensional systems, the formalism in Sec. \ref{1d_LC_construction:level2} can of course be extended, but the set of equations do not guarantee a limit cycle at $H=E_0$ in higher dimensions \cite{Ebeling2005}. We present a simple way of creating limit cycles in two dimensions and highlight its features under noise.

\section{\label{2d_LC:level1}Limit Cycles in Two Dimensional Systems}
In 1D, we have shown the presence of uncertainty in the  conjugates $(x,p_x)$. However, we had to introduce a nonlinear dissipation in momentum to obtain the limit cycle. This in turn produced a nonequilibrium bath. The uncertainty relation however, is not a consequence of the nonequilibrium bath. We illustrate this for a particle experiencing a central potential $V(r)$ in two dimension, in presence of a usual equilibrium/Langevin bath.

\subsection{\label{2d_LC_centrlpot:level2} Particle in a Central Potential}
 Consider the Hamiltonian in two dimension, with coordinates $(r,\phi)$ and momenta $(p_r, p_\phi),$
\begin{equation}
    H=\dfrac{p_r^2}{2}+\dfrac{p_\phi^2}{2r^2}+V(r).
\end{equation}
Since $\phi$ is cyclic, the angular momentum $p_\phi=\ell$  is conserved. Following \cite{Singh2025}, we use a radial Langevin bath which respects this conservation law. The equations of motion are now given by
\begin{equation}
    \dot{r}=p_r; \dot{p}_r=-\pdv{V}{r}+\dfrac{\ell^2}{r^3}-\gamma p_r+\sqrt{\gamma T}\xi(t); \dot{\phi}=\dfrac{\ell}{r^2}.
    \label{eq:2dLangevin}
\end{equation}

In the absence of noise ($T=0$) and $\ell\neq0$, the radial equation produces stable (unstable) limit cycle in $x$-$y$ plane  at $r=\sqrt{x^2+y^2}=r_c,$ where   
$r_c$ is the minimum (maximum) of the effective radial potential, $\widetilde{V}(r)=\ell^2/2r^2+V(r).$ In effect, stable or unstable limit cycles are also produced in $p_x$-$p_y$, $x$-$p_x$ and $y$-$p_y$ planes, 
\begin{align}
    p_x^2+p_y^2=\frac{\ell^2}{r_c^2};\ \frac{x^2}{r_c^2}+\frac{r_c^2p_x^2}{\ell^2}=1; \ \frac{y^2}{r_c^2}+\frac{r_c^2p_y^2}{\ell^2}=1. \label{eq:LCtraj_2D}
\end{align}

For any $T\ne0,$ the Fokker-Planck equation corresponding to the Langevin dynamics  \eqref{eq:2dLangevin} is
\begin{equation}
    \pdv{P}{t}=[H,P]+\pdv{p_r}(\gamma p_rP+\gamma T\pdv{P}{p_r})
\end{equation}
and admits a steady state given by the Boltzmann distribution $P(r,p_r,p_\phi=\ell)=\frac{e^{-\beta H}}{\mathcal{Z}_\ell(\beta)}$, where the partition function is
\begin{equation}
    \mathcal{Z}_\ell(\beta)=2\pi\sqrt{\dfrac{2\pi}{\beta}}\int_{0}^{\infty}\dd{r}\exp(-\beta\qty(\dfrac{\ell^2}{2r^2}+V(r))).\label{conditioncons}
\end{equation}
Of course, $\mathcal{Z}_\ell(\beta)$ must be finite, as required for a valid probability density function \cite{supp}. Provided $\ev{r^2}$ exists, these systems exhibit the position-momenta uncertainty relations
\begin{equation}
    \Delta x\Delta p_x=\Delta y\Delta p_y\geq\dfrac{\abs{\ell}}{2} \label{uncertaintycons}
\end{equation}
with the equality valid in the limit $T\to0$ \cite{Singh2025}. From the above relations, it is clear the there are positive lower bounds in each of the quantities, $\Delta x$, $\Delta y$ and $\Delta p_{x,y}$. 

We calculate the radial moments at small $T$ using the saddle-point approximation \cite{supp} and find that in the limit of $T\to 0$, $\ev{r}  = r_c$  and   $\ev{r^n} =\ev{r}^n\ \forall n\geq0.$  Thus, certainly the particle remains on the limit cycle as $ T\to0,$ having a residual energy $\ev{E}=E_c.$  Fluctuations in $r$ and $E$ vanish as $T\to0$, but the uncertainty relation (\ref{uncertaintycons}) holds.

\subsection{\label{2d_LC_r_alpha:level2}Example I: Power Law Potentials}
As an example, let us  consider the potentials
\begin{equation}
    V(r)=\frac{kr^\alpha}{\alpha};\quad k,\alpha>0
\end{equation}
which yield the integral in \eqref{conditioncons} finite. Note that, limit cycles are generated at $T=0$ even for potentials with $\alpha\in(-2,0)$ but under noise, the resulting steady state PDF does not converge on integration over all space. Thus, we restrict ourselves to $\alpha>0$ case. 

In the absence of noise ($T=0$) the particle is attracted to a limit cycle trajectory in the $x$-$y$ plane having radius $r_c$ and energy $E_c,$ given by
\begin{equation}
    r_c=\qty(\dfrac{\ell^2}{k})^{\frac{1}{\alpha+2}}\qcomma E_c= k \qty(\frac{2+\alpha}{2\alpha})r_c^\alpha.
\end{equation}
 The corresponding limit cycle trajectories in $p_x$-$p_y$, $x$-$p_x$ and $y$-$p_y$ planes are given by Eq. \eqref{eq:LCtraj_2D}.

\begin{figure}[t]
    \centering
    \includegraphics[width=\linewidth]{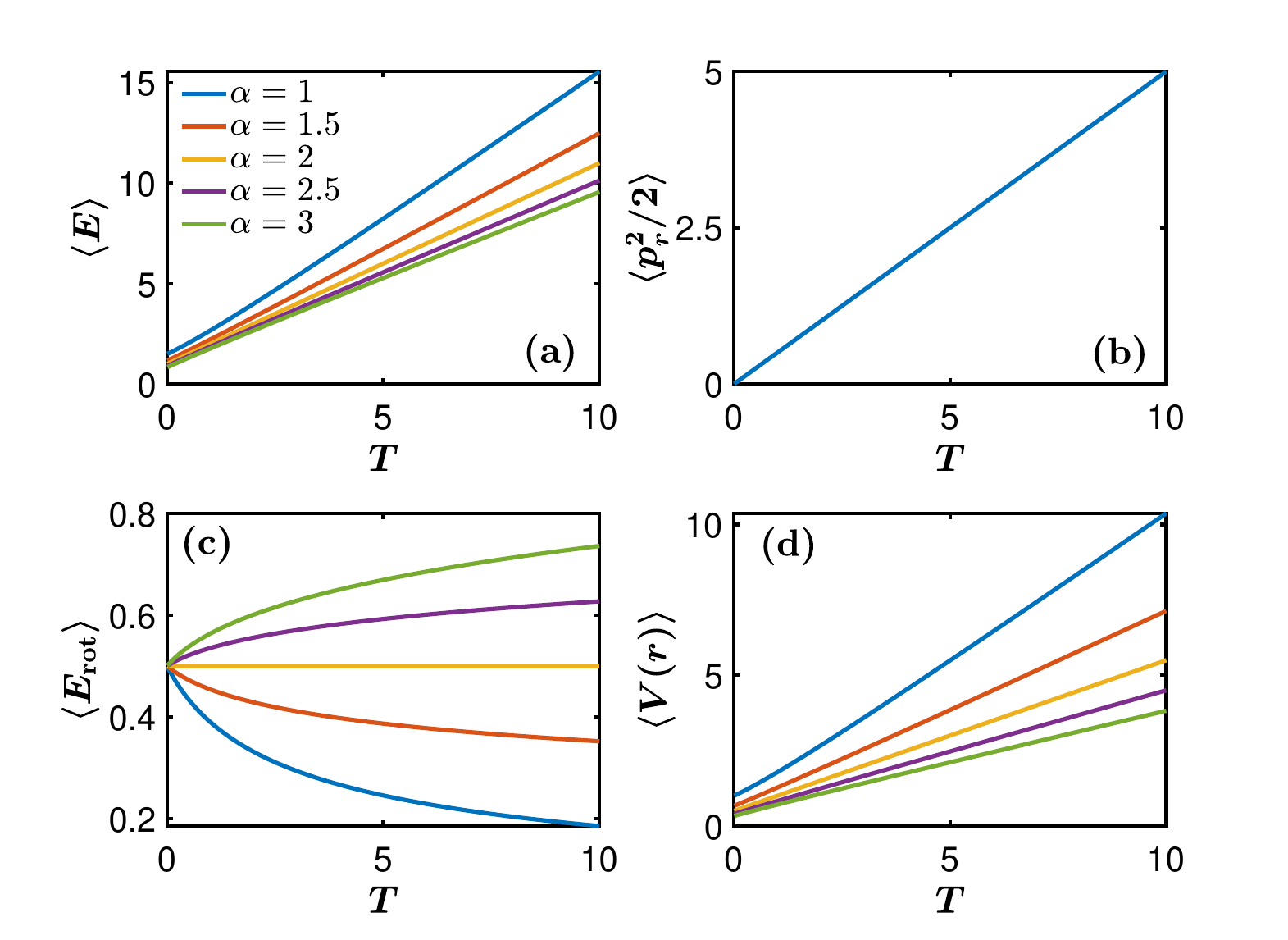}
    \caption{Variation of average energy, $\ev{E}=\ev{p_r^2/2}+\ev{E_{\mathrm{rot}}}+\ev{V(r)}$ with $T$ for $\alpha=1,1.5,2,2.5,3$. We choose $k=\ell=1$, hence, $r_c=1$. (a) shows the increase of $\ev{E}$ with $T$. The growth is faster for smaller $\alpha$ and $\ev{E}(T\to0)=E_c=(2+\alpha)/2\alpha$. (b) $\ev{p_r^2/2}=T/2$ for all values of $\alpha$. (c) $\ev{E_{\mathrm{rot}}}$ decreases for $\alpha<2$, is constant for $\alpha=2$ and increases for $\alpha>2$. (d) $\ev{V(r)}$ always increases with $T$ with a more rapid increase for smaller values of $\alpha$. Exact expressions are given in \cite{supp}.}
    \label{fig:energy_alpha}
\end{figure}

Apart from the uncertainty bound, these systems also exhibit interesting behaviour in the variation of average energy with temperature. $\ev{p_r^2/2}=T/2$ for all $\alpha$ and $\ev{E}$ is an increasing function of $T$ with $\ev{E}(T\to0)=E_c$ as expected. However, the behaviour of average rotational kinetic energy, $\ev{E_\mathrm{rot}}=\ev{\ell^2/2r^2}$ is quite different. It is a decreasing function of $T$ vanishing as $T\to\infty$ for $\alpha\in(0,2)$, a constant with value equal to $\frac{1}{2}\sqrt{k}\abs{\ell}$ for $\alpha=2$ while for $\alpha>2$, it increases with $T$ going to infinity as $T\to\infty$. Nevertheless, we have $\ev{E_\mathrm{rot}}(T\to0)=\ell^2/2r_c^2$ for all $\alpha$. Figure \ref{fig:energy_alpha} shows the variation of average energy with $T$ for different $\alpha$.

\subsection{\label{2d_LC_log}Example II: Logarithmic Potential}
 Another interesting example is the case of the logarithmic potential,
\begin{equation}
    V(r)=k\ln(\frac{r}{r_0});\quad k,r_0>0.
\end{equation}
As we shall see, this rather simple functional form of $V(r)$ reveals a temperature dependent existence of steady state as well as radial moments. 

At $T=0$, the above potential has a limit cycle at $r_c$ and energy $E_c$, given by
\begin{equation}
    r_c=\qty(\frac{\ell^2}{k})^{1/2}\qcomma E_c=\frac{k}{2}\qty(1+\ln(\frac{\ell^2}{kr_0^2})).
\end{equation}

For $T\neq0$, the Fokker-Planck equation has a steady state $\mathcal{Z_\ell(\beta)}^{-1}e^{-\beta H}$ with %only when $k\beta>1$, as
\begin{equation}
    \mathcal{Z}_\ell(\beta)= \qty(r_0\sqrt{\dfrac{2}{\beta}})^{k\beta}\pi^{3/2}\abs{\ell}^{1-k\beta}\Gamma\qty(\dfrac{k\beta-1}{2})
\end{equation}
only for $k\beta>1$ as required by the convergence of \eqref{conditioncons}. Thus, the line $T=k$ in the $(T,k)$ plane demarcates the region where we have a steady state from the region where we have no steady state $(T\geq k)$.

The $n$-th radial moment exists only when $k\beta>n+1$ given by 
\begin{equation}
    \ev{r^n}=\qty(\dfrac{\beta}{2})^{n/2}\abs{\ell}^n\frac{\Gamma\qty(\frac{k\beta-n-1}{2})}{\Gamma\qty(\frac{k\beta-1}{2})}.
\end{equation}
Thus, for $T\geq k/3$, $\ev{r^2}$, and consequently $\ev{x^2}=\ev{y^2}$, do not exist (even if the steady state exists) and we cannot talk about uncertainties. The uncertainty relation \eqref{uncertaintycons} definitively holds for $T<k/3$. As expected, $\Delta r$ and $\Delta E$ vanish  in the limit $T\to0$.

\section{\label{sec:dynamical}Dynamical Uncertainty}
All the inequalities we talked about in this article hold in the steady state. Therefore, we must have, dynamically, a time scale  $t^*$ such that the uncertainty relations become valid for all times $t>t^*$.

The  system evolves deterministically at $T=0,$  where  the  position and momenta of  the particle  at  any  time  $t$  depend on  the initial condition; such deterministic trajectories   result  in  $\Delta x(t)=0= \Delta p_x(t).$ One can, however,  choose an ensemble of particles  with   initial condition distributed according to some arbitrary PDF  to  get a  non-zero  variance  of $x$  (and  $y, p_{x,y}$) that  depends on  the choice of initial PDF. The situation for     
$T\ne0$  is different. Non-zero $T$  introduces  noise   to the dynamics, leading  the system  to a unique steady state,  independent of initial conditions.  

Naturally,  in the initial state, all degrees of  freedom  $x,y,p_x,p_y$ are  uncorrelated; thus,    $\Delta x\Delta p_x$  can be made  arbitrarily small at $t=0.$ We consider a fixed initial  value $(x,y,p_x,p_y) =(x_0,y_0,p_{x0},p_{y0})$ so that  $\Delta x=0=\Delta p_x$  at $t=0$ and follow the dynamics to  find out   how  $\Delta x\Delta p_x$   increases   as  time progresses  and  crosses the uncertainty bound $|\ell|/2$  at  $t=t^*$ before the system reaches its steady state. For  small $T,$ the steady state  value of $\Delta x\Delta p_x$ is slightly higher than   $|\ell|/2$ and thus one expects $t^* \sim \tau$, the time of relaxation to steady state.

\begin{figure}[t] \centering
\includegraphics[width=7.8cm]{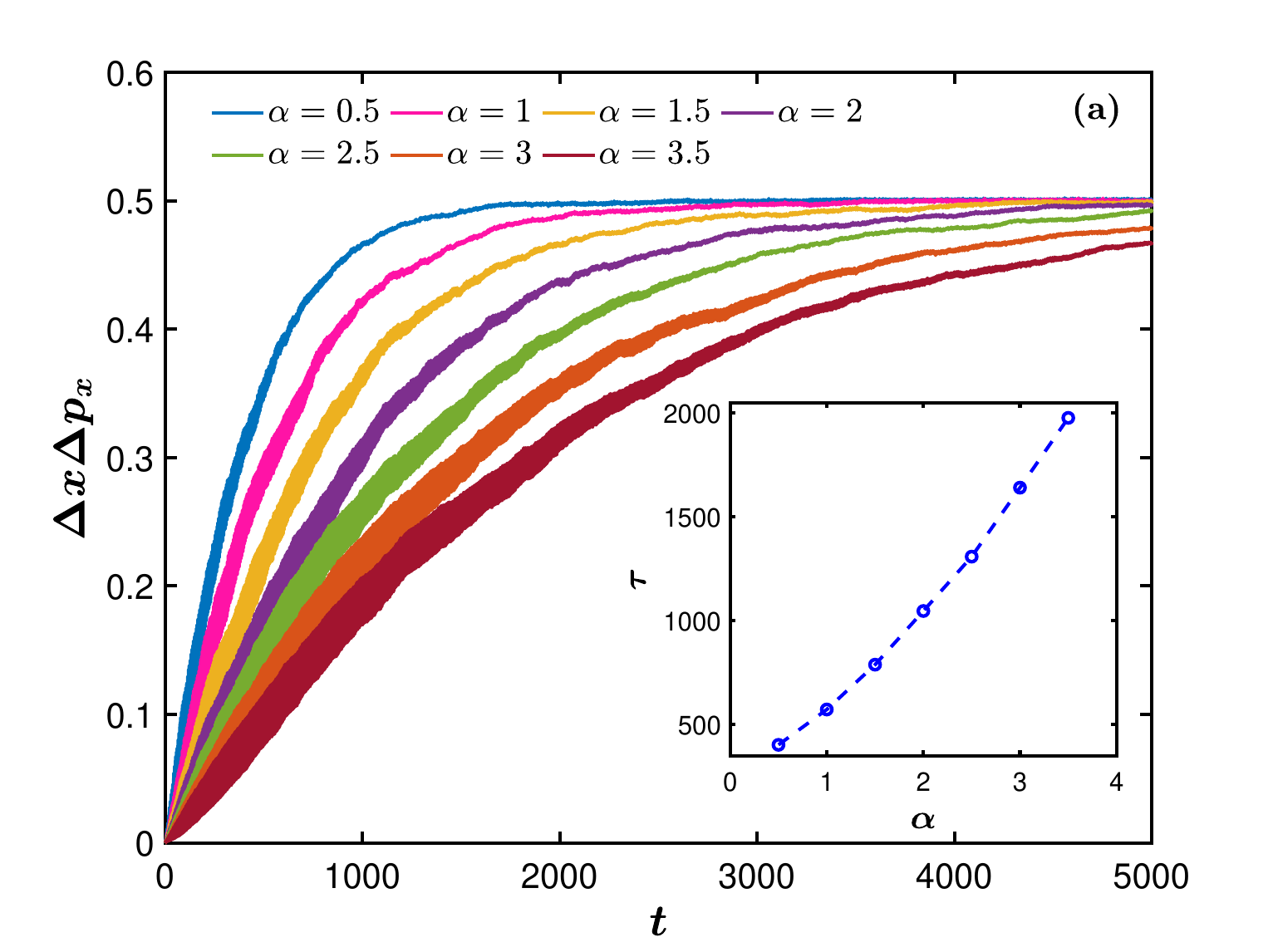}
\includegraphics[width=7cm]{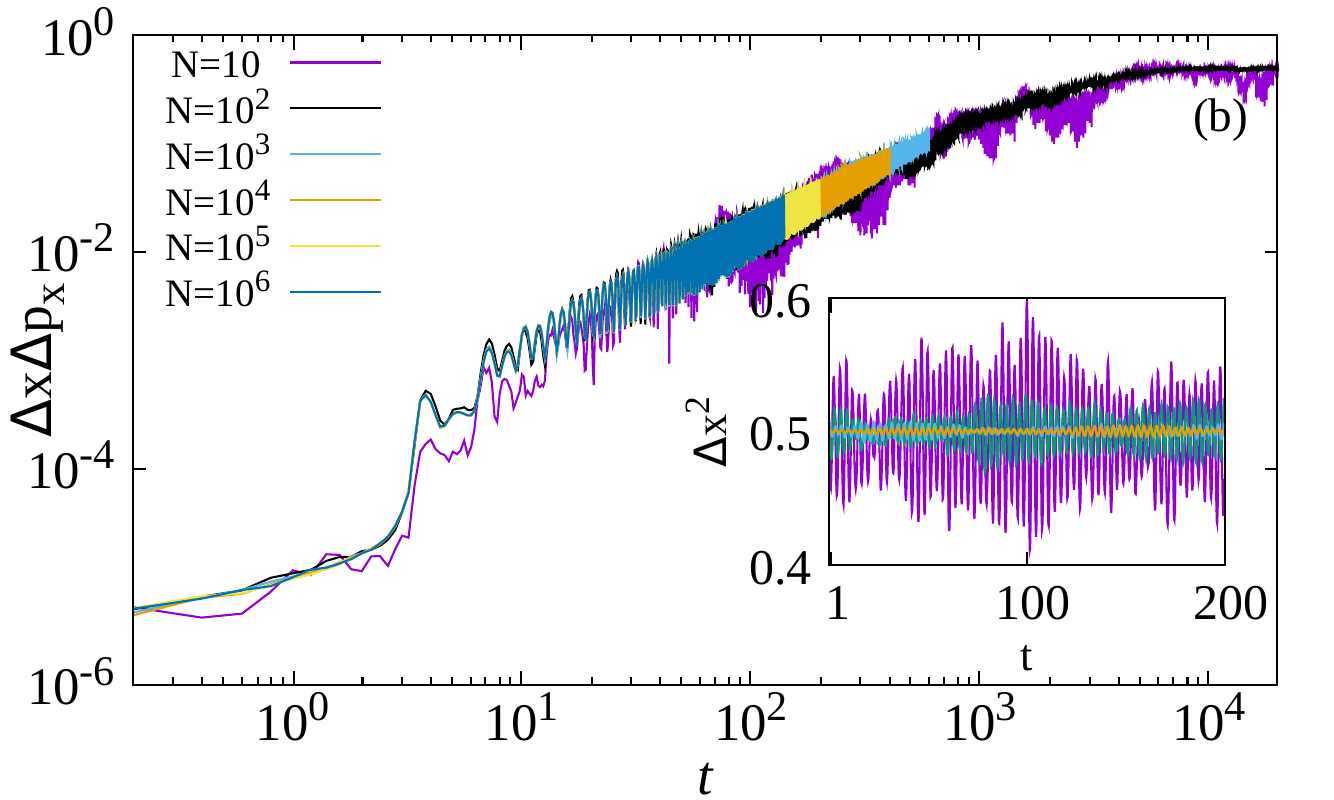}
\includegraphics[width=7cm]{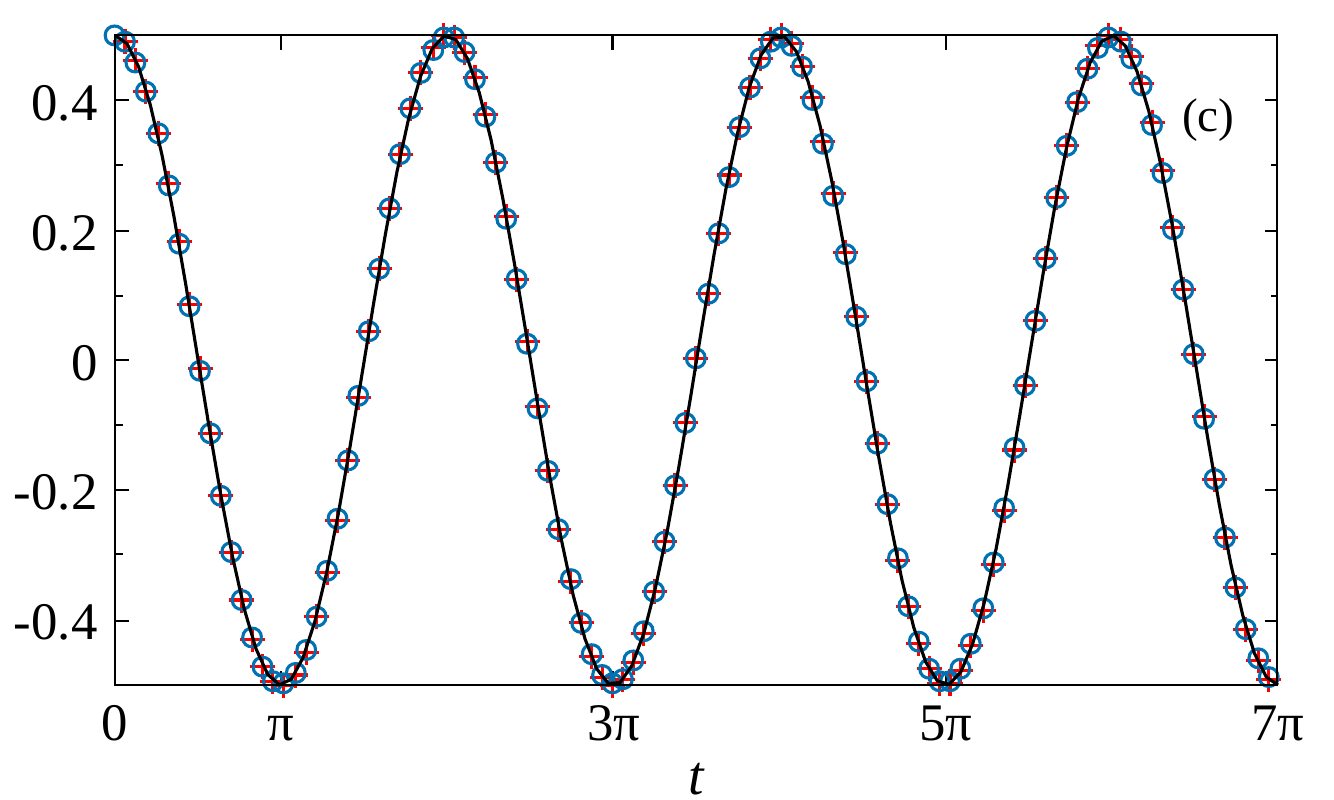}
\caption{(a) Variation of $\Delta x\Delta p_x$ with $t$ for $\alpha=0.5,1,1.5,2,2.5,3,3.5$, $\gamma=2$, $k=1=\ell$ at $T=10^{-3}$. The inset shows the relaxation time $\tau$ as a function of $\alpha$. (b)  Evolution of  $\Delta x \Delta p_x,$ when averaged over $N$ samples.  The large oscillations observed at small $t$ are  characteristic features  of the dynamics; they are robust, and  do not disappear even  after averaging over $N=10^6$ samples.  The inset  shows  steady state  oscillations; the amplitude of the oscillations  reduces as  the number of samples increase.    
(c)  The steady state average of  the autocorrelation functions $\ev{x(0)x(t)}$  (dots) and $\ev{y(0)y(t)}$ (crosses)  are compared with  $\cos(\omega t).$ $T=10^{-3},\alpha=2, \gamma=k=\ell=1$. Thus, $r_c=1$ and $\omega= \ell/r_c^2=1$.}
\label{fig:dyn}
\end{figure}

In  Fig. \ref{fig:dyn}(a) we show the evolution of $\Delta x\Delta p_x$  for a particle experiencing a central potential  $V(r) = \frac{k r^\alpha}\alpha$. We  keep  the  temperature very low, $T=10^{-3},$ and study the evolution  for different $\alpha$. With an  initial condition having angular momentum $\ell=1,$ one expects  $\Delta x\Delta p_x\ge\frac{1}{2}.$ It is evident from  Fig. \ref{fig:dyn}(a) that,    $\Delta x\Delta p_x$  approaches    this bound  as $t \to \infty.$   Initially at small $t$, the curves  appear  thicker - this  might give  an impression   that  the  data is not averaged enough. However,  for  such a low temperature,   the   trajectory  of the particle  in $x$-$p_x$  plane is  not far away from the limit cycle, and  most observables exhibit  sustained oscillations. As illustrated in  Fig. \ref{fig:dyn}(b) for $\alpha=2,$  the oscillations are in fact the property of the system and not a numerical artifact of inadequate ensemble averaging.   These  oscillations do survive  in the  steady state.   Their amplitude however  decreases   as  one  increase the  sample size $N$; this  is shown  in the inset of Fig.   \ref{fig:dyn}(b).  From the log-scale plot of  $\Delta x\Delta p_x - \frac{|\ell|}2$   as a  function of $t$  we obtain  the  relaxation  time $\tau;$  a plot of  $\tau$  vs $\alpha$ is shown in the inset of  Fig. \ref{fig:dyn}(a). For larger $\alpha,$ one may naively expect   the system to relax faster as the  potential-confinement is tighter. However,  we find that  the relaxation time  increases with $\alpha,$ owing to the inertia of the  particle.  For larger $\alpha,$  the  particle  lands  near the LC  with a  high  velocity  and goes  farther away  to the other side of the LC, creating 
oscillations, which   survive longer for larger $\alpha$. In fact,  the dissipation strength $\gamma$ controls how fast the system reaches  near  the LC $r=r_c.$ 

A  linear stability analysis about $r_c$ (at $T=0$) shows that the relaxation time of $r$ and $p_r$  is  $\tau_r = 2/\gamma;$ we find the time scale for relaxation  of $\ev{r}$ and $\ev{p_r}$  to be $\simeq \tau_r$ when $T$  is small \cite{supp}.  On the other hand, oscillations one  sees in $\ev{x},\ev{y}$ and $\ev{p_{x,y}}$  are  controlled by the temperature $T$ which  could not be captured from the linear stability analysis   at $T=0,$ reconfirming  the fact that   the system at any $T,$ however small it may be, cannot be treated as a perturbation  to the deterministic  dynamics  at $T=0.$

The oscillatory nature of the steady state at finite $T$  can also  be seen from the autocorrelation functions $C_x(t)=\ev{x(0)x(t)}$  and  $C_y(t)=\ev{y(0)y(t)},$  which are shown in Fig. \ref{fig:dyn}(c) for  $\alpha=2$ (SHO). We find that 
$C_x(t)= \cos{t}= C_y(t)$ both having a period of $2\pi,$  which  can be explained  as follows.   Since $\ev{r}$ relaxes very quickly to $r_c,$ we  can safely assume $r=r_c$  in the steady state. Thus $x(t)\sim r_c \cos(\omega t +\phi)$ and  $y(t)\sim r_c \sin(\omega t +\phi),$    with  initial position  of the particle $(x(0), y(0)) =  ( r_c \cos\phi, r_c \sin\phi).$  When  averaged over the initial condition, 
we get \be
\ev{x(0)x(t)}=\frac{r_c^2}{2\pi} \int_0^{2\pi} \dd{\phi} \cos(\omega t +\phi) \cos(\phi) = \frac{r_c^2}{2} \cos(\omega t) \nonumber   
\ee
which is  same as  $\ev{y(0)y(t)}$, as observed in Fig. \ref{fig:dyn}(c).

\section{\label{sec:discussion}Discussion}
\subsection{\label{1d_LC_nolimitcycle:level2} Uncertainties without Limit Cycle}
Though limit cycle systems undoubtedly show uncertainties in observables defining the limit cycle phase plane, they are not limited  to these systems; similar features can also be  observed  in systems which exhibit restricted noisy motion  in a plane,  around a close curve, but do not  reduce to a limit cycle system   when noise is turned off. We illustrate this using two examples: a 1D system having a continuum of fixed points which form a closed curve and the case of an active Brownian particle (ABP) in 2D harmonic confinement.

\subsubsection{System  with  a Continuum of Fixed Points}
As illustrated in Sec. \ref{2d_LC:level1}, the uncertainty relations are not a consequence of the nonequilibrium bath. In Sec. \ref{1d_LC_construction:level2}, we encountered a steady state, non-Boltzmann in $H$. However, these nonequilibrium steady states can be equilibrium steady states of another dynamics which may not necessarily generate limit cycles at $T=0$ but they may behave like a limit cycle system when $T\neq0$. These systems are expected to show uncertainty relations. We illustrate this by devising an alternate dynamics which will generate the same steady state as \eqref{eq:1d_steadystate} but may not produce a limit cycle.
\begin{figure}[t]
    \centering
   \includegraphics[width=\linewidth]{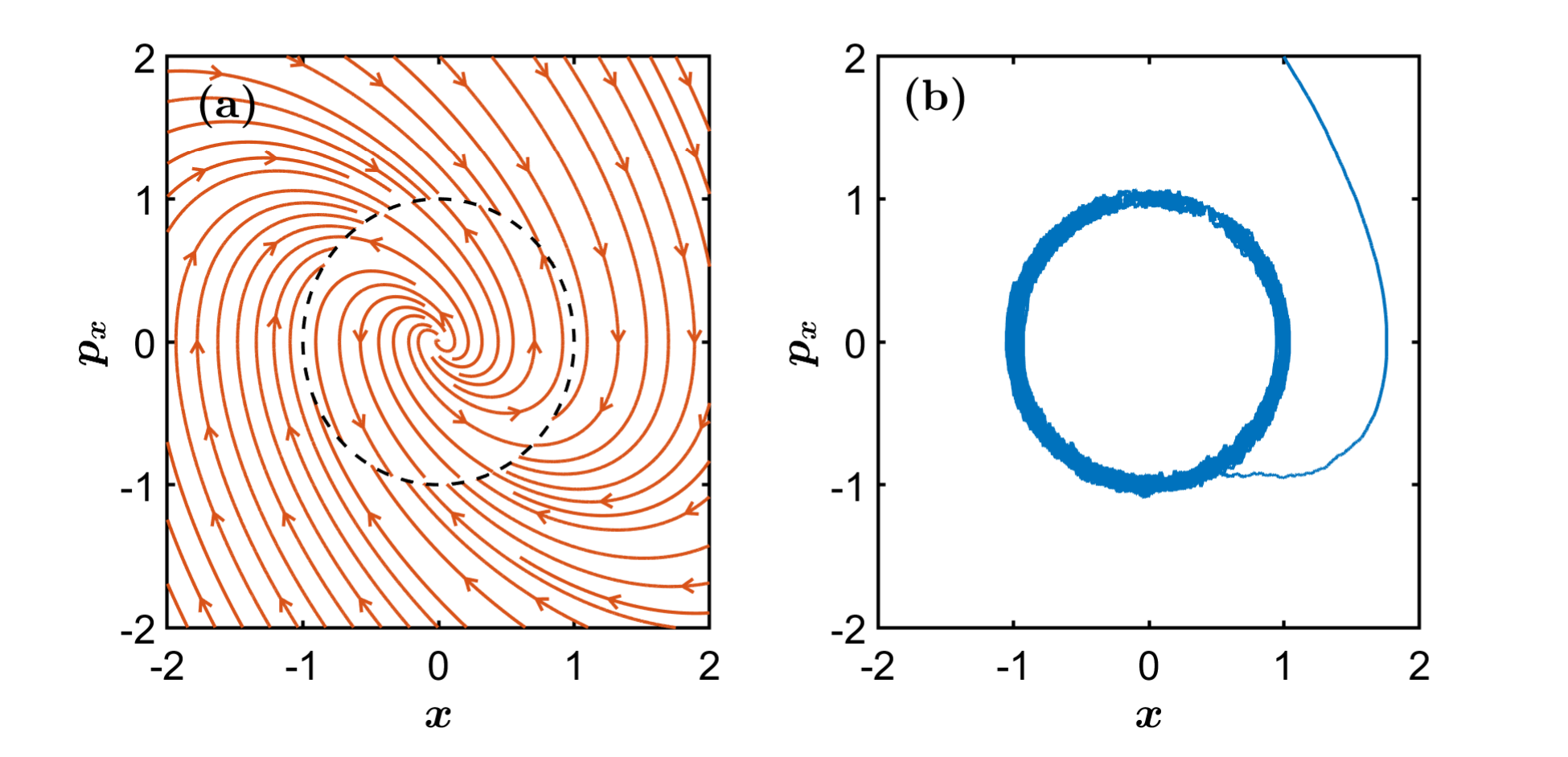}
    \caption{(a) Flow generated by the system of equations \eqref{eq:LCunderT} for $E_0=0.5,\gamma=1=k$. The ellipse (here, circle) of fixed points $H=E_0$ is shown by dashed lines (b) A sample path of the dynamics \eqref{eq:LCunderT} at $T=10^{-3}$ with $\gamma=1=k$ and $E_0=0.5$.}
    \label{fig:LCunderT_vectorplot}
\end{figure}

We define an auxiliary Hamiltonian, $\widetilde{H}$, corresponding to $H$ in \eqref{eq:1d_steadystate} as
\begin{equation}
    \widetilde{H}=\int\dd{H}g(H).
\end{equation}
Consider the dynamics \cite{supp}
\begin{equation}
     \dot{x}=\pdv{\widetilde{H}}{p_x} \qcomma \dot{p}_x=-\pdv{\widetilde{H}}{x}-\gamma\pdv{\widetilde{H}}{p_x}+\sqrt{\gamma T}\xi(t). \label{eq:1d_nocycle}
\end{equation}
The Fokker-Planck equation for the above dynamics reads as
\begin{equation}
    \pdv{P}{t}=\qty[\widetilde{H},P]+\pdv{p_x}(\gamma\pdv{\widetilde{H}}{p_x}P+\gamma T\pdv{P}{p_x}),
\end{equation}
which leads to the steady state distribution 
\begin{equation}
    P(x,p_x)=\dfrac{1}{\mathcal{Z}}\exp(-\beta\widetilde{H}(x,p_x)). \label{eq:1d_ss_nolimcycle}
\end{equation}
Here, $\mathcal{Z}$ is given by Eq. \eqref{eq:partitionfunction_1D}. We proceed with $H=p_x^2/2+kx^2/2$, the Hamiltonian of a 1D SHO considered in Sec. \ref{1d_LC_sho:level2} with 
\begin{equation}
    \widetilde{H}(x,p_x)=\dfrac{(H(x,p_x)-E_0)^2}{2E_0}. \label{eq:auxiliaryH_sho}
\end{equation}
The dynamics according to \eqref{eq:1d_nocycle} is 
\begin{equation}
\begin{aligned}
    \dot{x}&=\dfrac{p_x}{E_0}(H-E_0),\\ \dot{p}_x&=-\qty(\dfrac{H-E_0}{E_0})\qty(kx+\gamma p_x)+\sqrt{\gamma T}\xi(t). \label{eq:LCunderT}
\end{aligned}
\end{equation}

At $T=0$, the above system has an ellipse of fixed points at $H=E_0$ shown in Fig. \ref{fig:LCunderT_vectorplot} (a) and produces no limit cycle. Nevertheless, at non-zero $T$, it is described by the steady state \eqref{eq:1d_ss_nolimcycle} with $\widetilde{H}$ as defined in \eqref{eq:auxiliaryH_sho} and the uncertainty relation \eqref{eq:1d_uncertainty} holds.

\begin{figure}[t]
    \centering
   \includegraphics[width=\linewidth]{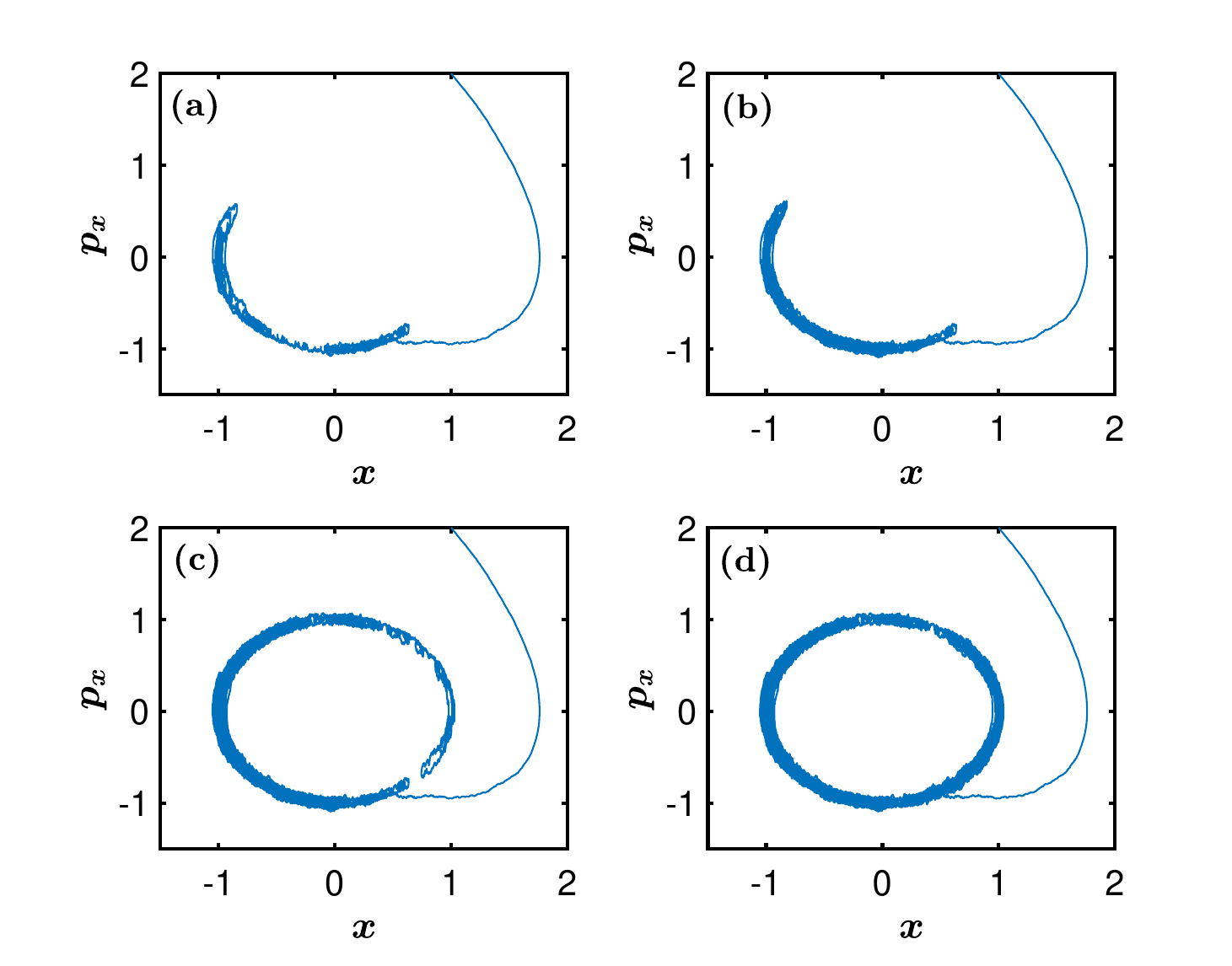}
    \caption{The particle starts from the initial condition $(1,2)$ at $T=10^{-3}$ with $\gamma=k=1$ and $E_0=0.5$ (a) Snapshot of the trajectory at $t=400$. The particle has moved clockwise around the ellipse of fixed points (b) Snapshot at $t=1400$. The particle has moved back anticlockwise over the ellipse of fixed points as indicated by the increased line density of plot. (c) Snapshot at $t=3800$. The particle has moved clockwise again, almost covering completely the ellipse of fixed points. (d) Snapshot at $t=5000$. The particle has covered fully the ellipse of fixed points, however, the increased line density in the plot is a signature of the particle moving repeatedly clockwise as well as anticlockwise over the ellipse of fixed points.}
    \label{fig:LCunderT_sample}
\end{figure}

As can be seen in Fig. \ref{fig:LCunderT_vectorplot} (b), under weak noise, the particle moves over the ellipse of fixed points as expected from the steady state distribution. However, the motion is quite different from a limit cycle system where the noisy trajectories are definitively clockwise or anticlockwise depending on the direction of rotation of the particle in the limit cycle at $T=0$. In the system considered here, the noisy trajectories are about the ellipse of fixed points but have no directional preference. This is illustrated in Fig. \ref{fig:LCunderT_sample}. 

\begin{figure}[t]
    \centering 
   \includegraphics[width=\linewidth]{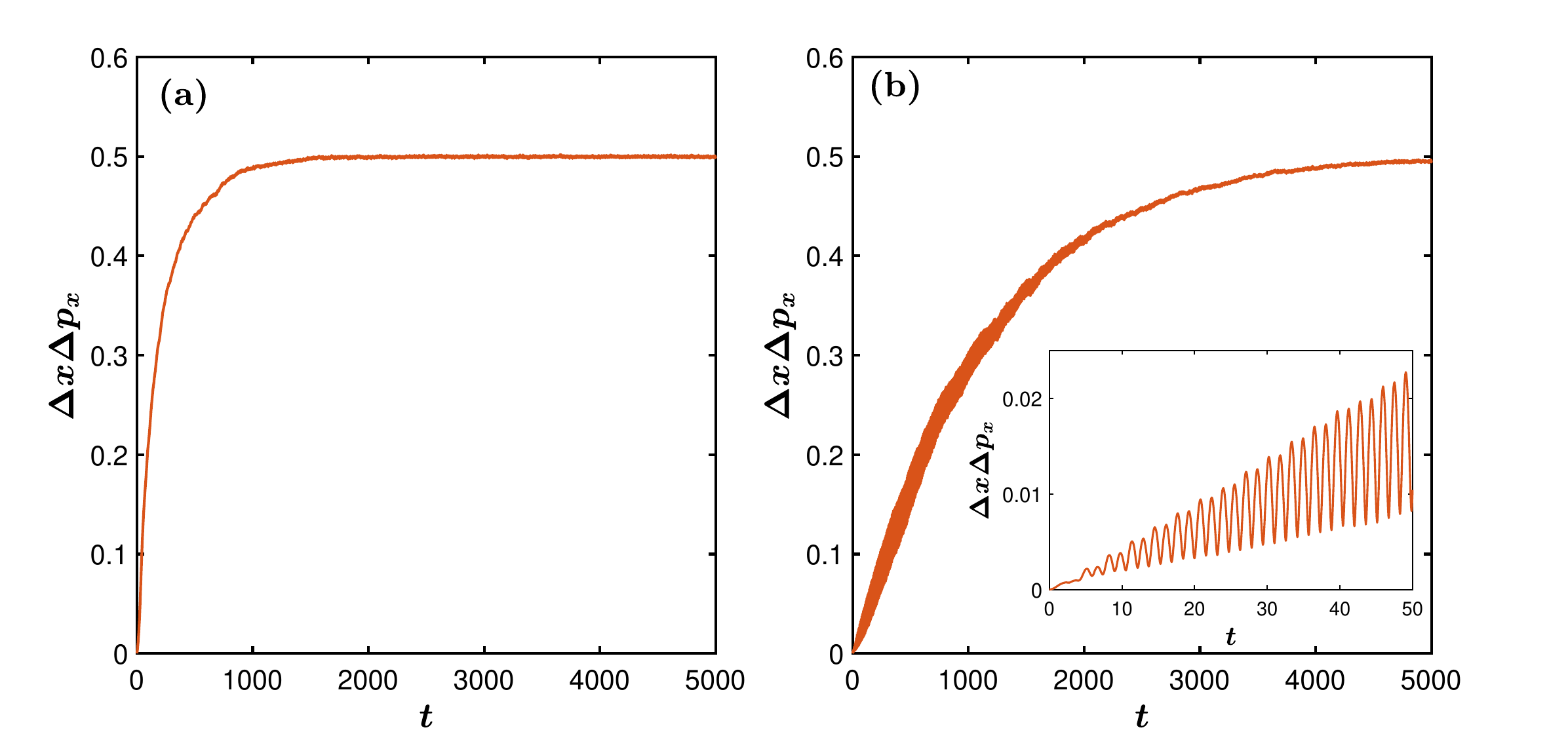} 
    \caption{(a) and (b) show $\Delta x\Delta p_x$ as a function of $t$ for dynamics \eqref{eq:LCunderT} and \eqref{eq:sho1d} respectively. The simulations have been performed at $T=10^{-3}$ for $\gamma=k=1$ and $E_0=0.5$. At low noise, the uncertainties roughly hold the equality in the uncertainty relation \eqref{eq:1d_uncertainty} and approach $E_0/\omega_0=0.5$. In (b), the system has a limit cycle at $T=0$ and hence, the oscillatory behaviour shown in the inset as opposed to (a).}
    \label{fig:1D_unc}
\end{figure}

As a result, the oscillatory behaviour in the position-momentum moments as well as the position-momentum uncertainty highlighted in Sec. \ref{sec:dynamical} is not exhibited in this system. 
The $(\Delta x\Delta p_x)(t)$ generated by dynamics \eqref{eq:LCunderT} is compared with the dynamics given in Eq. \eqref{eq:sho1d} for $\gamma=k=1$ at $T=10^{-3}$ in Fig. \ref{fig:1D_unc}.

\subsubsection{\label{sec:ABP}Active Brownian Particle in a Harmonic Trap}
ABP in harmonic confinement in 2D  \cite{Basu2019, Malakar2020, Chaudhuri2021} is another system which does not exhibit limit cycle behaviour but can show non-trivial uncertainty bounds in the observables, $x$ and $y$, which define the plane of motion of the particle. We consider an ABP with translational diffusion constant $D_t$ and rotational diffusion constant $D_r$ in harmonic confinement $V(r)=kr^2/2$,  with internal degree of freedom $\theta$, governed by the equations
\begin{equation}
\begin{aligned}
    \dot{x}&=v_0\cos\theta-\mu kx+\sqrt{D_t}\xi_1(t),\\
    \dot{y}&=v_0\sin\theta-\mu ky+\sqrt{D_t}\xi_2(t),\\
    \dot{\theta}&=\sqrt{D_r}\xi_3(t).
\end{aligned}
\end{equation}
where as before, $\xi_i(t)$ are Gaussian white noise with $\ev{\xi_i(t)}=0$ and $\ev{\xi_i(t)\xi_j(t')}=2\delta_{ij}\delta(t-t')$. The translational mobility is denoted by $\mu$ and the particle self-propels with speed $v_0$ in the direction $(\cos\theta,\sin\theta)$. Following \cite{Malakar2020,Chaudhuri2021}, we have in the steady state,
\begin{equation}
\begin{aligned}
    &\ev{x}=\ev{y}=0,\\
    &\ev{x^2}=\ev{y^2}=\frac{\ev{r^2}}{2}=\frac{D_t}{\mu k}+\frac{v_0^2}{2\mu k (D_r+\mu k)}.
\end{aligned}
\end{equation}

We look at different regions of the $(D_t,D_r)$ parameter space where we can get non-trivial bounds on $\Delta x,\Delta y$. When the particle experiences only rotational diffusion $(D_t=0,D_r\neq0)$, we have
\begin{equation}
    0\leq\Delta x=\Delta y\leq \frac{v_0}{\sqrt{2}\mu k}.
\end{equation}
where the equality in the upper bound holds in the limit $D_r\to0$ while the equality in the lower bound holds as $D_r\to\infty$. Here, the nature of the bounds is quite different from what we have been analyzing so far; $\Delta x$ and $\Delta y$ have a non-trivial upper bound, i.e., the uncertainty in $x$ and $y$ cannot exceed a certain maximum value.

We recover the familiar nature of uncertainty, i.e., non-trivial lower bounds, in the regimes of $D_r=0$ with $D_t\neq0$ and $D_t=\lambda D_r=D$, $\lambda$ being an appropriate dimensionful constant. When $D_r=0$ with $D_t\neq0$, 
\begin{equation}
    \Delta x=\Delta y\geq\frac{v_0}{\sqrt{2}\mu k}
\end{equation}
where the equality holds in the limit $D_t\to0$. In the case of $D_t=\lambda D_r=D$, we obtain
\begin{equation}
    \Delta x=\Delta y\geq \begin{cases}
        \frac{v_0}{\sqrt{2}\mu k}, & v_0\leq\sqrt{2\lambda}\mu k\\ \\
        \sqrt{\frac{\sqrt{2\lambda} v_0}{\mu k}-\lambda}, &v_0>\sqrt{2\lambda}\mu k
    \end{cases}
\end{equation}
where the equality holds for $D\to0$ when $v_0\leq\sqrt{2\lambda}\mu k$ whereas for $D=-\mu k\lambda+v_0\sqrt{\frac{\lambda}{2}}$ when $v_0>\sqrt{2\lambda}\mu k$.

\subsection{\label{subsec:Discussion_ss} Origin of uncertainty bound}
It is true, at least mathematically, that the uncertainty relations are simply features of some specific probability distributions.   If a  stochastic variable $z$  has a normal distribution  with  mean $z_c$ and standard deviation $\sigma,$  we can always construct  new   variables $x= z \cos\phi$ and  $y= z \sin\phi,$   with  $\phi$ distributed uniformly in the range $[0,2\pi)$ which  leads to $\Delta x \Delta y \ge \frac{z_c^2}{2}.$  Such a distribution, however, cannot generate   uncertainty bound  on  conjugate  variables  $(x, p_x= -\frac1z \sin\phi),$ $(y, p_y= \frac1z \cos \phi).$  A  two dimensional SHO   at finite temperature   does exhibit  noisy motion about  a limit cycle $r=r_c$ in the $x$-$y$ plane  where  the marginal  distribution $P(x,y)$ is peaked. However,  $P(x,y)$  is far from a trivial   normal distribution; this feature creates bounds on  $\Delta x, \Delta y$  along with  bounds on  $\Delta p_{x,y}.$ It is worth appreciating that these distributions are created naturally as steady states of limit cycle systems under noise, an important and a ubiquitous physical system. As illustrated in Sec. \ref{1d_LC_nolimitcycle:level2}, the uncertainties are not only limited to pure limit cycle systems (systems exhibiting limit cycle oscillations at $T=0$) but also applicable to systems with an entirely different dynamics mimicking limit cycle systems under noise.
%{\clr\sout{, that is, sharing the same steady state.}, that is, fluctuating about some closed curve even in the limit of vanishing noise.}

It should be noted that the validity of the uncertainty relation \eqref{uncertaintycons} hinges on the existence of $\ev{r^2}$ moment. For the dynamics defined in \eqref{eq:2dLangevin}, $\exp(-\beta H)$ always solves the steady state Fokker-Planck equation but this probability density integrated over all space may not necessarily converge. The situation becomes even more intricate when there is temperature dependent convergence of PDF as well as moments. The logarithmic potential, an apparently simple form which shows such behaviour, was considered in Sec. \ref{2d_LC_log}. The $r^\alpha$ potentials defined in Sec. \ref{2d_LC_r_alpha:level2} are free from such complications and always admit a steady state while upholding the uncertainty relations \eqref{uncertaintycons}.

\section{\label{sec:conclusion}Conclusion}
In conclusion, we  show  that whenever  a particle makes  a noisy trajectory around a  closed  curve,  even in the limit of vanishing noise, the   observables  that  define  the    phase plane  would exhibit an uncertainty  relation, i.e., the  product of the standard deviations  of these  observables  must be bounded from below by a  positive constant. The simplest example  of such a system is  a limit cycle system   subjected to thermal noise.   In one dimension, we use the framework of canonical dissipative systems to construct limit cycles in the  $x$-$p_x$  plane controlled by parameter $E_0$ which carries the dimension of energy and show that  for SHO with frequency $\omega_0$, $\Delta x \Delta p_x \ge  E_0/\omega_0$; note  that  $E_0/\omega_0$  is  the simplest construct based on  the parameters of the model  which  has the dimension of angular momentum, same as the dimension of $\Delta x \Delta p_x.$   In two dimension, the dynamics of   particles in a  central potential does not  alter the angular momentum $\ell$ of the system and  and an added radial dissipation  brings the  particle  to follow a circular trajectory in  real space ($x$-$y$  plane) - in effect, elliptical limit cycles are also generated  in  $x$-$p_x$ and $y$-$p_y$ plane. Further when  the  radial coordinate  experiences  a   stochastic force,   the system reach a steady state  having  a  Boltzmann distribution wrt a temperature that  obeys  fluctuation-dissipation relation; the angular momentum of the system, however, remains conserved. In  these systems,  we find  a  position momenta uncertainty relation in  the equilibrium state:  $\Delta x \Delta p_x \ge  \abs{\ell}/2,$  irrespective of the  nature of the  radial potential  (so long as $\ev{r^{2}}$ exists). 

Such a thermodynamic bound   on fluctuations of observables defining a phase plane  is  not  limited to limit cycle systems. We show that  systems   that produce a continuum of fixed points  forming a closed curve  in the $x$-$p_x$ plane,  also exhibit  the same when  noise is added; we show it  explicitly for a one dimensional system, with   Hamiltonian  $H =\frac{1}{2E_0} \left(\frac{1}{2} p_x^2 + \frac{1}{2}\omega_0^2  x^2 - E_0 \right)^2,$ which resulted in $\Delta x \Delta p_x \ge  E_0/\omega_0.$ 
We obtain similar bounds on an ABP in a 2D harmonic trap in the $x$-$y$ plane.

We believe, the  theoretical predictions  of the  model can be verified in  systems  where  particles move in a  noisy trajectory around a closed curve. One such system is  of course  the active  Brownian particle   in a  harmonic trap  in two dimension \cite{Takatori2016, Basu2019, Dauchot2019, Malakar2020, Chaudhuri2021, Caraglio2022, Nakul2023} considered in Sec. \ref{sec:ABP}. 
%{\clm (remove this line?)- for very high motility they  move  stochastically  around a  circle, far away from the  minimum of the potential.}  
Another example is the  chiral active particles, which  naturally form     noisy  trajectories around   a circle \cite{Kmmel2013, tenHagen2014, Mano2017, Caprini2019}.

%\bibliographystyle{apsrev4-2} % Tell bibtex which bibliography style to use
%\bibliography{references.bib}
%

\end{document}

\title{Supplemental Material for ``Statistical Uncertainties  of Limit Cycle Systems in Langevin Bath"}
\author{Dipesh K. Singh} \email{dks20ms176@iiserkol.ac.in} \author{P. K. Mohanty}
\email{pkmohanty@iiserkol.ac.in}
\affiliation {Department of Physical Sciences, Indian Institute of Science Education and Research Kolkata, Mohanpur, 741246, India.}
\date{\today}

\begin{abstract}
In this Supplemental Material, we give a  description of the Fokker-Planck equation  rewritten in a specific form  (in the main text)  along with its method of solution. We also show explicit calculation of partition function for $1$D limit cycle system under noise constructed out of SHO.  We  also  provide   the   exact expressions of the mean  energy for $r^\alpha$ potentials. We elaborate further on the dynamical behaviour of $r^\alpha$ potentials aided by linear stability analysis at $T=0$.
\end{abstract}
\maketitle

\section{Fokker-Planck Equation}
For a set of $N$-variable Langevin equations of the form 
\begin{equation}
    \dot{\zeta}_i=h_i(\zeta,t)+\sum_{j}g_{ij}(\zeta,t)\xi_j(t),
\end{equation}
where $\xi_i(t)$ are Gaussian white noise with $\ev{\xi_i(t)}=0$ and $\ev{\xi_i(t)\xi_j(t')}=2\delta_{ij}\delta(t-t')$, the Fokker-Planck equation for the probability density $P(\zeta,t)$ is given by
\begin{equation}
    \pdv{P}{t}=-\sum_{i}\pdv{\zeta_i}(D_iP)+\sum_{i,j}\frac{\partial^2}{\partial \zeta_i\partial \zeta_j}(D_{ij}P),
\end{equation}
where $\displaystyle D_i=h_i+\sum_{k,j}\pdv{g_{ij}}{\zeta_k}g_{kj}$ and $D_{ij}=\displaystyle\sum_{k}g_{ik}g_{jk}$ are the so-called drift and diffusion coefficients respectively \cite{Risken1996}.

For a noisy dynamics derived from the dynamics generated by a Hamiltonian $H$, as is our case, with the equations
\begin{equation}
    \dot{x}_i=\pdv{H}{p_i}\qcomma \dot{p}_i=-\pdv{H}{x_i}-f_i(x,p)+g_i(x,p)\xi_i(t)
\end{equation}
with $\ev{\xi_i(t)}=0$ and $\ev{\xi_i(t)\xi_j(t')}=2\delta_{ij}\delta(t-t')$, the non-trivial drift and diffusion coefficients are
\begin{equation}
    D_{x_i}=\pdv{H}{p_i}\qcomma D_{p_i}=-\pdv{H}{x_i}-f_{i}+g_{i}\pdv{g_{i}}{p_i} \qand
    D_{p_i p_i}=g_{i}^2
\end{equation}
using which the Fokker-Planck equation for probability density $P(x,p,t)$ can be written as 
\begin{equation}
    \pdv{P}{t}=[H,P]+\sum_{i}\pdv{p_i}(f_i P-g_{i}\pdv{g_{i}}{p_i}P+\pdv{p_i}(g_i^2 P)),
\end{equation}
where $[A,B]=\displaystyle\sum_{i}\qty(\pdv{A}{x_i}\pdv{B}{p_i}-\pdv{A}{p_i}\pdv{B}{x_i})$ is the Poisson bracket of $A$ and $B$. For the special case when $g_{i}$ is independent of $p_i$, the Fokker-Planck equation becomes
\begin{equation}
    \pdv{P}{t}=[H,P]+\mathcal{L}P=[H,P]+\sum_{i}\pdv{p_i}(f_i P+g_i^2\pdv{P}{p_i}).
\end{equation}

Corresponding to Eq. (1) in the main text, the steady state Fokker-Planck equation is given by
\begin{equation}
    [H,P]+\pdv{p_x}(\gamma g(H)\pdv{H}{p_x}P+\gamma T\pdv{P}{p_x})=0.
\end{equation}
As mentioned in the main text, the above is solved using the ansatz $P=P(H)$. With this assumption, the Poisson bracket vanishes and using chain rule and the problem simplifies to solving $\mathcal{L}P=0$, i.e.,
\begin{equation}
    \pdv{p_x}(\gamma g(H)\pdv{H}{p_x}P+\gamma T\pdv{P}{H}\pdv{H}{p_x})=0\imp \pdv{p_x}(\gamma T\pdv{H}{p_x}\qty(\beta g(H)P+\pdv{P}{H}))=0.
\end{equation}
The above reduces to solving the equation
\begin{equation}
    \pdv{P}{H}+\beta g(H)P=0 
\end{equation}
which of course has a solution given by 
\begin{equation}
    P(H)=\frac{1}{\mathcal{Z}}\exp(-\beta\int\dd{H}g(H))
\end{equation}
where the partition function $\mathcal{Z}$ should be finite for $P$ to be a valid probability density function. 

Similarly, corresponding to Langevin equations 
\begin{equation}
    \dot{r}=p_r\qcomma \dot{p}_r=-\pdv{V}{r}+\frac{\ell^2}{r^3}-\gamma p_r+\sqrt{\gamma T}\xi(t)\qcomma \dot{\phi}=\frac{\ell}{r^2}, \label{eq:radialLangevin}
\end{equation}
the steady state Fokker-Planck is given by
\begin{equation}
    [H,P]+\pdv{p_r}(\gamma p_rP+\gamma T\pdv{P}{p_r})=0.
\end{equation}
Using the ansatz $P=P(H)$, and noting that $\pdv{H}{p_r}=p_r$, we have
\begin{equation}
    \pdv{p_r}(\gamma p_r P+\gamma T\pdv{P}{H}\pdv{H}{p_r})=\pdv{p_r}(\gamma p_r\qty( P+ T\pdv{P}{H}))=0.
\end{equation}
Thus, $P(H)\sim \exp(-\beta H)$ solves the above equation. We demand that the probability density integrated over all space converges and hence, the partition function
\begin{equation}
    \mathcal{Z}_\ell(\beta)=\int_{0}^{2\pi}\dd{\phi}\int_{-\infty}^{\infty}\dd{p_r}\int_{0}^{\infty}\dd{r}P(r,p_r)=\int_{0}^{2\pi}\dd{\phi}\int_{-\infty}^{\infty}\dd{p_r}\int_{0}^{\infty}\dd{r}\exp(-\beta H)=\mathrm{finite}.
\end{equation}
If above holds, $P(H)=\frac{e^{-\beta H}}{\mathcal{Z}_\ell(\beta)}$ is the steady state solution.

In Sec. V A of the main text, we wanted to devise a dynamics which would yield $P\sim\exp(-\beta\widetilde{H})$ as the steady state. To do so, consider the set of equations
\begin{equation}
    \dot{x}=\pdv{\widetilde{H}}{p_x}\qcomma \dot{p}_x=-\pdv{\widetilde{H}}{x}-f+\sqrt{\gamma T}\xi(t)
\end{equation}
where $f$ is undetermined as of now. The steady state Fokker-Planck equation for above is given by
\begin{equation}
    [\widetilde{H},P]+\pdv{p_x}(fP+\gamma T\pdv{P}{p_x})=0.
\end{equation}
Using the ansatz $P=P(\widetilde{H})$, the above simplifies to 
\begin{equation}
    \pdv{p_x}(fP+\gamma T\pdv{P}{\widetilde{H}}\pdv{\widetilde H}{p_x})=0 \imp \pdv{p_x}(\gamma \pdv{\widetilde{H}}{p_x}\qty(\frac{f}{\gamma\pdv{\widetilde{H}}{p_x}} P+T\pdv{P}{\widetilde{H}}))=0.
\end{equation}
Choose $f=\gamma\pdv{\widetilde{H}}{p_x}$. The above equation reduces to  
\begin{equation}
    \pdv{p_x}(\gamma\pdv{\widetilde{H}}{p_x}\qty(P+T\pdv{P}{\widetilde{H}}))=0
\end{equation}
which leads to the solution $P(H)\sim\exp(-\beta\widetilde{H})$.

\section{Calculation of Partition Function}
We compute the partition function (8) in the main text. We need to evaluate the integral
\begin{equation}
    \mathcal{Z}_{E_0}(\beta)=\int_{-\infty}^{\infty}\dd{p_x}\int_{-\infty}^{\infty}\dd{x}\exp(-\dfrac{\beta}{2E_0}\qty(\dfrac{p_x^2}{2}+\dfrac{kx^2}{2}-E_0)^2).
\end{equation}
This is best done using a coordinate transform
\begin{equation}
    x=\sqrt{\dfrac{2}{k}}\ r\cos\phi  \qcomma p_x=\sqrt{2}\ r\sin\phi \imp \dd{x}\dd{y}=\dfrac{2r}{\sqrt{k}}\dd{r}\dd{\phi}.
\end{equation}
The partition function thus becomes
\begin{equation}
    \mathcal{Z}_{E_0}(\beta)=\dfrac{2}{\sqrt{k}}\int_0^{2\pi}\dd{\phi}\int_{0}^{\infty}\dd{r}r\exp(-\dfrac{\beta}{2 E_0}(r^2-E_0)^2)=\sqrt{\dfrac{2E_0}{k\beta}} \pi^{3/2}\qty(1+\mathrm{erf}\qty(\sqrt{\dfrac{\beta E_0}{2}})).
\end{equation}

\section{Calculation of Radial Moments for Steady States of Central Potentials}
For the distribution $e^{-\beta H}$ with $H=p_r^2/2+\ell^2/2r^2+V(r)$, the radial moments for $n\geq0$ are given by
\begin{equation}
    \ev{r^n}=\dfrac{\displaystyle\int_0^\infty\dd{r} r^n\ \exp(-\beta \widetilde{V}(r))}{\displaystyle\int_0^\infty\dd{r} \exp(-\beta \widetilde{V}(r))};\quad \widetilde{V}(r)=\dfrac{\ell^2}{2r^2}+V(r).
\end{equation}
$\widetilde{V}(r)$ is the effective radial potential as defined earlier. In the $T\to0$ limit, (equivalently, $\beta\to\infty$), the value of the integral is dominated by the strongest minimum. Let $r_c$ denote the radius of the limit cycle corresponding to this minimum. Performing the saddle-point approximation around $r=r_c$, the integral becomes
\begin{equation}
    \ev{r^n}\sim\dfrac{\displaystyle\int_0^\infty\dd{r} r^n\ \exp(- \frac{\beta\widetilde{V}''(r_c)}{2}(r-r_c)^2)}{\displaystyle\int_0^\infty\dd{r} \exp(- \frac{\beta\widetilde{V}''(r_c)}{2}(r-r_c)^2)}=\dfrac{\displaystyle\int_{-r_c}^\infty\dd{r} (r+r_c)^n\ \exp(- \frac{\beta\widetilde{V}''(r_c)r^2}{2})}{\displaystyle\int_{-r_c}^\infty\dd{r} \exp(- \frac{\beta\widetilde{V}''(r_c)r^2}{2})}.
\end{equation}
The above Gaussian can be integrated in terms of confluent hypergeometric function ${}_1F_{1}(a;b;z)$ and Gamma function. The integral evaluates to
\begin{equation}
    \ev{r^n}\sim\frac{b^{-\frac{n}{2}} \left(\Gamma\qty(\frac{n+1}{2}) \, _1F_1\left(-\frac{n}{2};\frac{1}{2};-r_c^2 b\right)+r_c n\sqrt{b}\Gamma\qty(\frac{n}{2}) \, _1F_1\left(\frac{1-n}{2};\frac{3}{2};-r_c^2 b\right)\right)}{\sqrt{\pi } \left(\text{erf}\left(r_c \sqrt{b}\right)+1\right)};\ b=\dfrac{\beta \widetilde{V}''(r_c)}{2}.
\end{equation}
The asymptotic form of ${}_1F_1(p;q;z)$ as $z\to-\infty$, when $\Gamma(q-p)$ is finite is given by
\begin{equation}
    {}_1F_1(p;q;z)\sim \dfrac{\Gamma(q)}{\Gamma(q-p)}(-z)^{-p}.
\end{equation}
Finally taking the limit $\beta\to\infty$ (equivalently, $b\to\infty$),
\begin{equation}
    \ev{r^n}\sim \dfrac{b^{-n/2}}{2\sqrt{\pi}}\qty(\dfrac{\Gamma\qty(\frac{n+1}{2})\Gamma\qty(\frac{1}{2})r_c^n\ b^{n/2}}{\Gamma\qty(\frac{n+1}{2})}+\dfrac{n\Gamma\qty(\frac{n}{2})\Gamma(\frac{3}{2})r_c^nb^{n/2}}{\Gamma\qty(\frac{n+2}{2})})
\end{equation}
which simplifies to 
\begin{equation}
    \ev{r^n}(\beta\to\infty)=r_c^n.
\end{equation}
Thus, the particle is definitely on $r=r_c$, the limit cycle trajectory, as $T\to0$ with both of $\Delta r(T\to0)=0$ and $\Delta E(T\to0)=0$.

\section{Linear Stability Analysis}
\begin{figure}[ht]
    \centering
    \includegraphics[width=\linewidth]{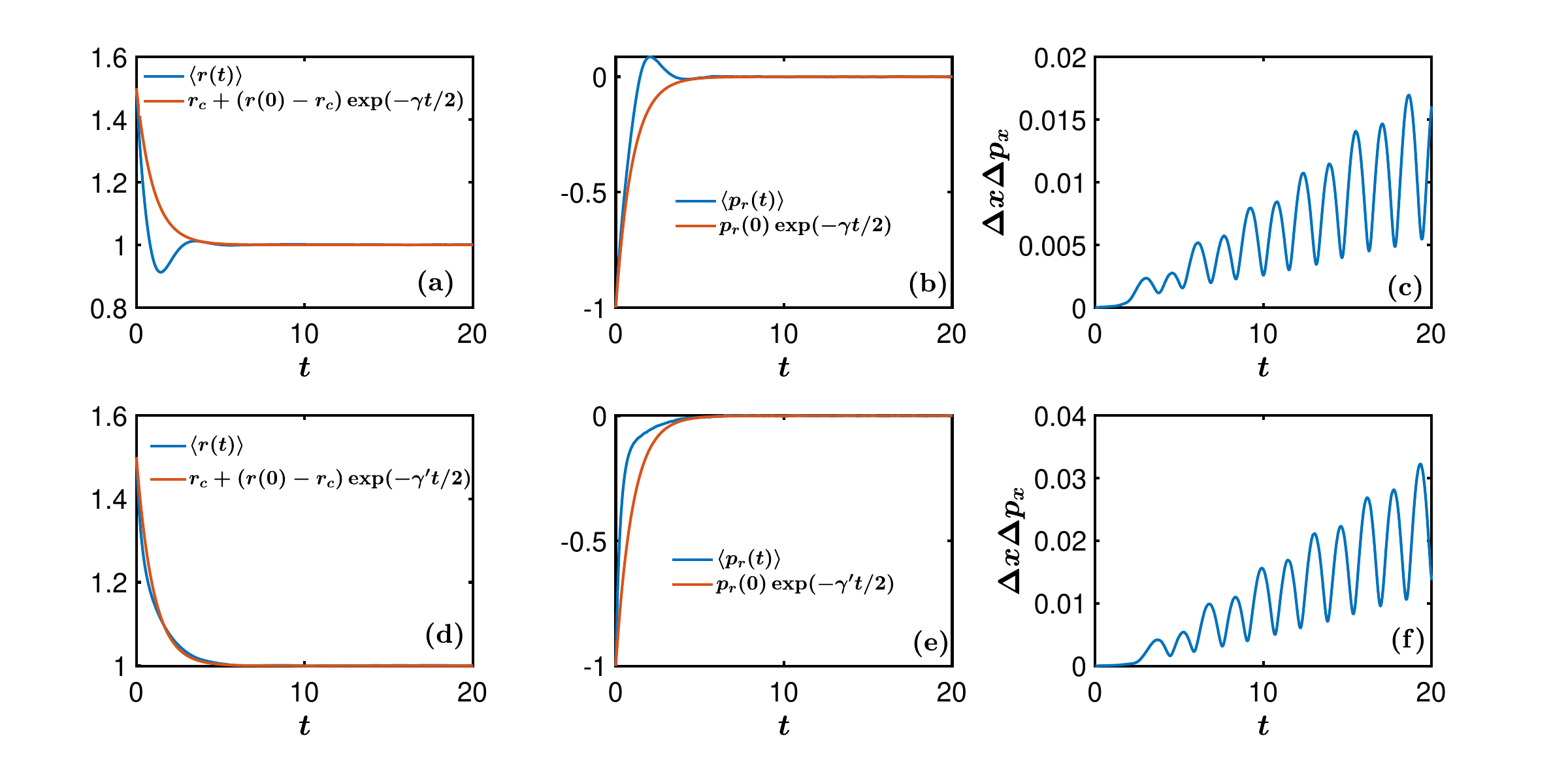}
    \caption{Simulations  for $V(r)=r$ at $T=10^{-3}$ with $\ell=1.$   (a), (b), (c):  here $\gamma=2$   which results in the oscillatory decay of $\ev{r(t)}$ and $\ev{p_r(t)}.$  The data  is compared with 
  the only  time scale $\gamma=2.$  (d), (e), (f): here $\gamma=4$ which results  in   an  exponential decay  with two time scales. The data   is compared  with     the dominating time-scale $\gamma'=2.$    In  both  cases,  however, the oscillatory behaviour  prevails in the evolution of  uncertainties, as can be seen in $\Delta x\Delta p_x$ in (c) and (f).}
    \label{fig:timescale_sup_alpha1}
\end{figure}

\begin{figure}[ht!]
    \centering
    \includegraphics[width=\linewidth]{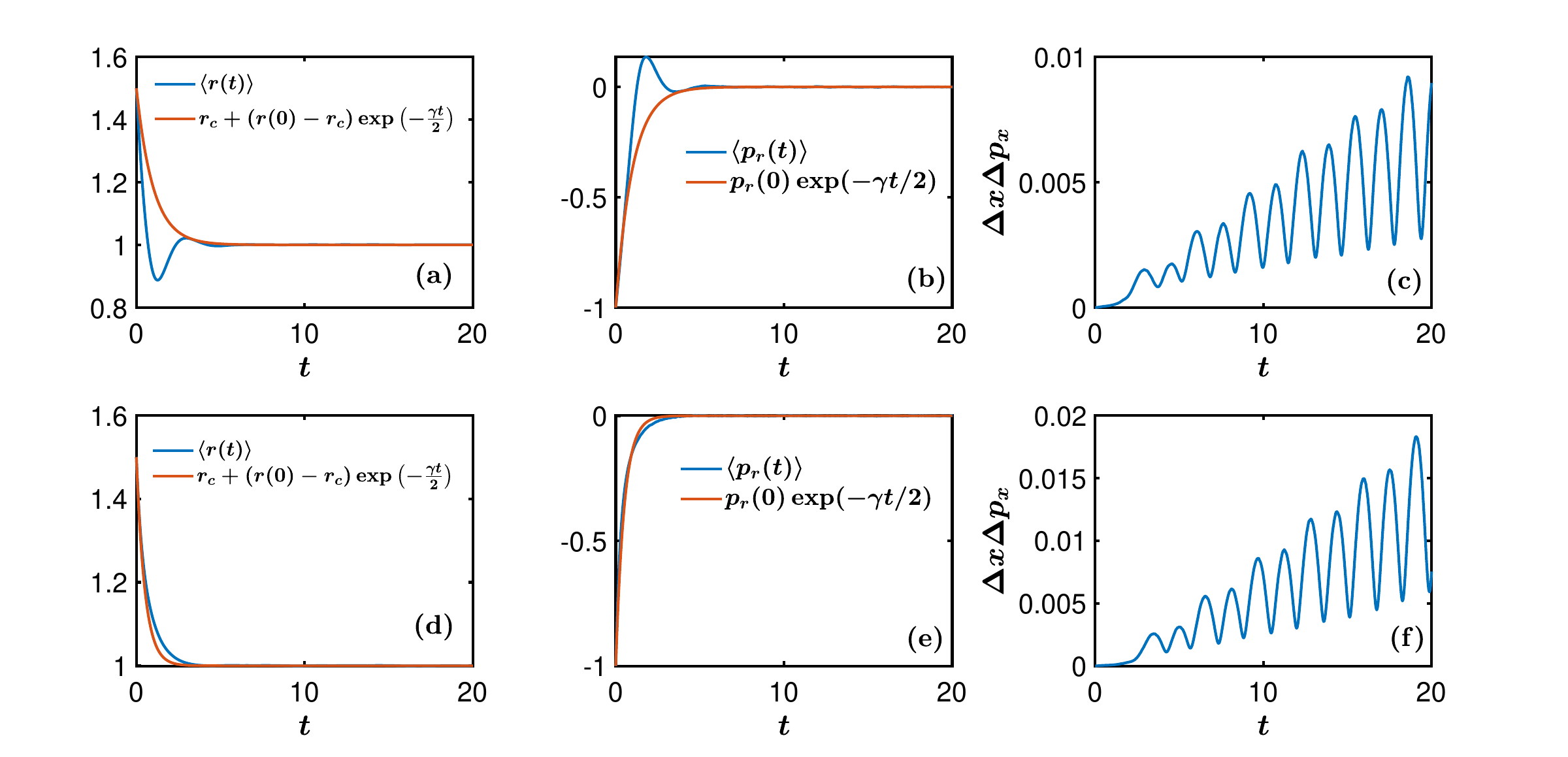}
    \caption{Simulations for $V(r)=r^2/2$ at $T=10^{-3}$ with $\ell=1$. (a),(b),(c):  here $\gamma=2$ 
 which  results in  oscillatory decay of $\ev{r(t)}$ and $\ev{p_r(t)}.$ (d), (e), (f):  here $\gamma=4$ which  shows exponential decay. This is  the marginal case where both eigenvalues are same.   
  %there is only one time scale $\gamma=2$. In (d), (e), (f) both eigenvalues are equal and we . We see the oscillatory decay of $\ev{r(t)}$ and $\ev{p_r(t)}$ in (a) and (b) while an exponential decay in (d) and (e) respectively, with an approximate time scale of $2/\gamma$ which can be obtained from calculation of eigenvalues. 
  In both cases, oscillatory behaviour still prevails in the uncertainties, as can be seen in $\Delta x\Delta p_x$ in (c) and (f).}
    \label{fig:timescale_sup}
\end{figure}

We perform linear stability analysis around the limit cycle generated by dynamics \eqref{eq:radialLangevin} at $T=0$ to get an idea about the time scale of decay of small perturbation from the limit cycle trajectory. The dynamics at $T=0$ is given by
\begin{equation}
    \dot{r}=p_r\qcomma \dot{p}_r=-\pdv{\widetilde{V}}{r}-\gamma p_r\qcomma \dot{\phi}=\frac{\ell}{r^2},
\end{equation}
where we have used the effective radial potential $\widetilde{V}(r)=\ell^2/2r^2+V(r)$. Consider small perturbations $\delta r,\delta p_r$ and $\delta\phi$ about the limit cycle at $r_c$. Considering terms only first order in these perturbations, we get
\begin{equation}
    \dv{t}\mqty(\delta r\\ \delta p_r\\ \delta\phi)=\mqty(0& &1 &&0\\-\widetilde{V}''(r_c)&& -\gamma&&0\\ -\frac{2\ell}{r_c^3}&&0&&0)\mqty(\delta r\\ \delta p_r\\ \delta\phi)
\end{equation}
where prime represents derivative wrt $r$. The eigenvalues and the corresponding eigenvectors of the above stability matrix are
\begin{equation}
    \lambda_0=0\qcomma\lambda_{\pm}=-\frac{\gamma}{2}\pm\frac{\sqrt{\gamma^2-4\widetilde{V}''(r_c)}}{2} \qq{with} \vb{v}_{\lambda_0}=\mqty(0\\0\\1)\qcomma \vb{v}_{\lambda_{\pm}}=\mqty(-\frac{r_c^3}{2\ell}\lambda_{\pm}\\ \\ -\frac{r_c^3}{2\ell}\lambda_{\pm}^2\\ \\1).
\end{equation}
We therefore have
\begin{equation}
   \mqty(\delta r\\ \delta p_r\\ \delta\phi)(t)=c_{0}\vb{v}_{\lambda_0}+c_{+}e^{t\lambda_{+}}\vb{v}_{\lambda_{+}}+c_{-}e^{t\lambda_{-}}\vb{v}_{\lambda_{-}}.
\end{equation}
In case of small damping, such that $\gamma^2-4\widetilde{V}''(r_c)<0$, we get an oscillatory decay of $\delta r(t)$ and $\delta p_r(t)$ to $0$ with a time scale $2/\gamma$. In case of large damping, such that $\gamma^2-4\widetilde{V}''(r_c)\ge0$, we have an exponential decay of $\delta r(t)$ and $\delta p_r(t)$ to $0$ with the time scales given by $\lambda_{\pm}^{-1}$. The time scale obtained is roughly the relaxation time of $\ev{r(t)}$ to steady state at small $T$ as shown in Fig. \ref{fig:timescale_sup_alpha1} and \ref{fig:timescale_sup}. In fact, the feature of oscillatory decay for smaller values of $\gamma$ and exponential decay for larger values of $\gamma$ according to the conditions mentioned here is also preserved.

The dynamics \eqref{eq:radialLangevin} at $T=0$ can be expressed in Cartesian coordinates using the relations
\begin{equation}
    x=r\cos\phi\qcomma y=r\sin\phi\qcomma p_x=p_r\cos\phi-\frac{\ell}{r}\sin\phi\qcomma p_y=p_r\sin\phi+\frac{\ell}{r}\cos\phi.
\end{equation}
Using above, we get
\begin{equation}
\begin{aligned}
    &\dot{x}=p_x\qcomma & &\dot{p}_x=-x\qty(\frac{V'(r)}{\sqrt{x^2+y^2}}+\gamma\frac{xp_x+yp_y}{x^2+y^2})\\
    &\dot{y}=p_y\qcomma & &\dot{p}_y=-y\qty(\frac{V'(r)}{\sqrt{x^2+y^2}}+\gamma\frac{xp_x+yp_y}{x^2+y^2})
\end{aligned}
\end{equation}
where prime denotes differentiation wrt $r$ and $r=\sqrt{x^2+y^2}$. Note that on the limit cycle, $r=r_c$ and $p_r=0$. We consider $V(r)=kr^\alpha/\alpha;\ \alpha>0$ potentials. For simplicity consider $\ell=k=1$ such that $r_c=1$. For this case, the linear stability analysis yields the following system of equations. We denote $\sin\phi$ as $s$ and $\cos\phi$ as $c$ respectively.
\begin{equation}
    \dv{t}\mqty(\delta x\\ \delta p_x\\ \delta y\\ \delta p_y)=\mqty(0 && 1 &&0 &&0\\ (\alpha -2) s^2 -\alpha +\gamma  sc+1&& -\gamma c^2 && -c((\alpha -2)s+\gamma c) && -\gamma cs\\ 0 &&0&&0&&1\\ s((2-\alpha )c+\gamma s) && -\gamma cs && -s((\alpha -2) s+\gamma c)-1 && -\gamma s^2)\mqty(\delta x\\ \delta p_x\\ \delta y\\ \delta p_y)
\end{equation}
The stability matrix above is a function of time owing to the time dependence of $\phi$. In fact, here, we have $\dot\phi=1$ so $\phi(t)=t+\phi_0$ on the limit cycle with $\phi_0$ being a constant fixed by the initial conditions. The above system of equations cannot be integrated analytically and hence, it is not possible to get some analytical form of time scale of decay of perturbations. If one considers a fixed $\phi$, the eigenvalues and eigenvectors will be different for different times and hence, the linear stability analysis does not shed any light on the behaviour of the system.  In fact, for some values of $\phi$ the eigenvalues indicate a diverging solution which is unphysical in context of a stable limit cycle. The same problems arise even when  we work with an averaged $\phi$ in the stability matrix. Thus, linear stability analysis which gave us reasonable estimate of the  relaxation of  $\ev{r(t)}, \ev{p_r(t)},$ fails to capture the oscillatory behaviour of $\ev{x(t)}$, $\ev{y(t)}$ and $\ev{p_{x,y}(t)}$ at low $T.$ 

\section{Dynamical Uncertainties}
\begin{figure}[t]
    \centering
    \includegraphics[width=0.5\linewidth]{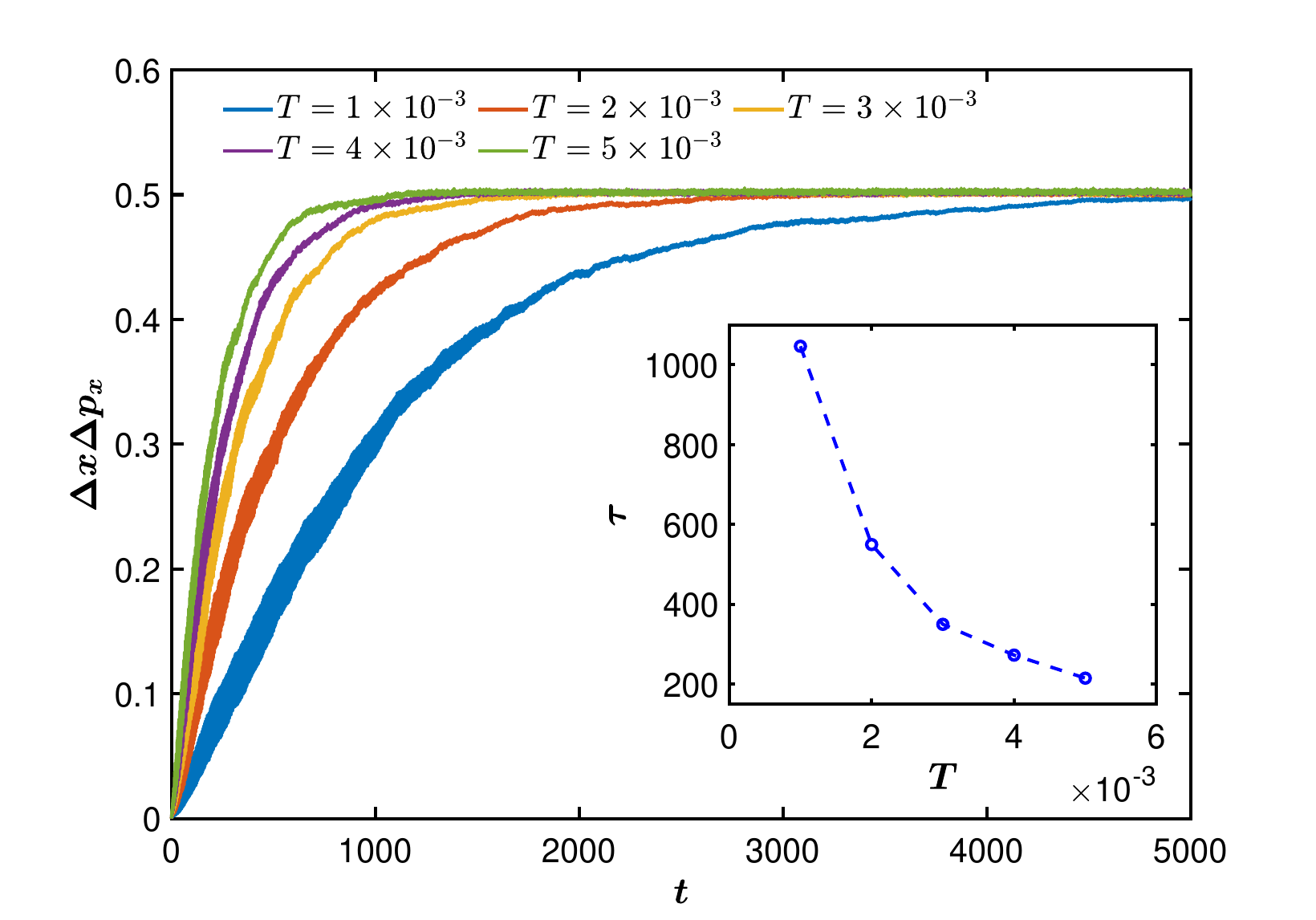}
    \caption{Variation of $\Delta x\Delta p_x$ with $t$ for $V(r)=r^2/2$ for $T=(1,2,3,4,5)\times10^{-3}$ $(\gamma=2$, $\ell=1)$. $\tau$ has been found out by fitting the curves to $\frac{1}{2}(1+T)(1-\exp(-t/\tau))$ where $\frac{1}{2}(1+T)$ is the steady state value of $\Delta x\Delta p_x$ for $\alpha=2$.}
    \label{fig:uncertainty_T}
\end{figure}

\begin{figure}[h]
    \centering
    \includegraphics[width=\linewidth]{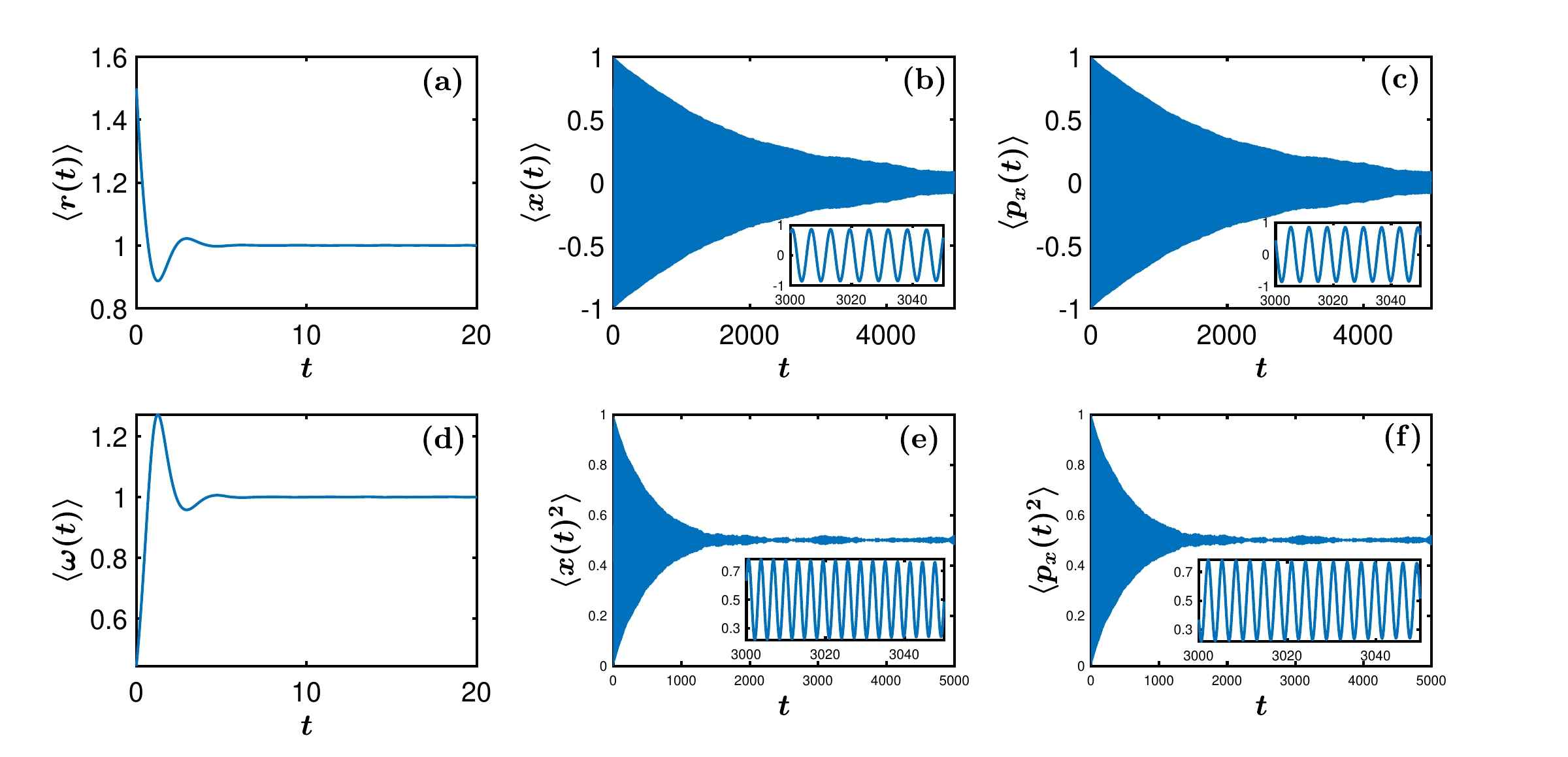}
    \caption{The above has been obtained from simulation of $V(r)=r^2/2$ at $T=10^{-3}$ with $\gamma=2$ and $\ell=1$. (a) and (d) show the quick relaxation of $\ev{r(t)}$ and $\ev{\dot{\phi}=\omega(t)}$ to steady state. On the other hand, $\ev{x(t)},\ev{p_x(t)},\ev{x(t)^2}$ and $\ev{p_x(t)^2}$ have a much longer relaxation time scale as shown in (b), (c), (e) and (f) respectively. The insets show the prevailing oscillatory behaviour of these moments.}
    \label{fig:sho_moments}
\end{figure}

We illustrate some additional features of the dynamics for $V(r)=kr^\alpha/\alpha$ potentials. As one would expect, the relaxation time, $\tau$ decreases with increasing $T$. We show this feature for $\alpha=2$ in Fig. \ref{fig:uncertainty_T}. $\ev{r(t)}$ relaxes much faster than any of $\ev{x(t)^n},\ev{y(t)^n},\ev{p_x(t)^n}$ and $\ev{p_y(t)^n}$ where $n\ge1$. This has been shown in Fig. \ref{fig:sho_moments} for $\alpha=2$.
%\clearpage
\section{Expressions of Average Energy  for power-law central potentials}
 The expressions for $\ev{E}$ and $\ev{E_{\mathrm{rot}}}$ for   $V(r)= \frac{k r^\alpha}{\alpha}$ are given below.

\begin{center}
    \begin{tabular}{|c|c|c|}
        \hline
         $\alpha$& $\ev{E_\mathrm{rot}}$ & $\ev{E}$\\ \hline
        & &\\
        $1$& $\dfrac{\beta ^2 k^2 \ell^2 G_{0,3}^{3,0}\left(\left.\frac{1}{8} k^2 \ell^2 \beta ^3\right|
\begin{array}{c}
 -\frac{1}{2},0,0 \\
\end{array}
\right)}{8 G_{0,3}^{3,0}\left(\left.\frac{1}{8} k^2 \ell^2 \beta ^3\right|
\begin{array}{c}
 0,\frac{1}{2},1 \\
\end{array}
\right)}$& 
$\dfrac{3}{8\beta}\qty(4+\dfrac{\beta ^3 k^2 l^2 G_{0,3}^{3,0}\left(\left.\frac{1}{8} k^2 l^2 \beta ^3\right|
\begin{array}{c}
 -\frac{1}{2},0,0 \\
\end{array}
\right)}{G_{0,3}^{3,0}\left(\left.\frac{1}{8} k^2 l^2 \beta ^3\right|
\begin{array}{c}
 0,\frac{1}{2},1 \\
\end{array}
\right)})$ \\ & &\\ \hline & &\\

        $1.5$& $\dfrac{3 G_{0,7}^{7,0}\left(\left.\frac{3.57k^4\ell^6\beta^7}{10^6}\right|
\begin{array}{c}
 0,\frac{1}{6},\frac{1}{4},\frac{1}{2},\frac{1}{2},\frac{3}{4},\frac{5}{6} \\
\end{array}
\right)}{\beta  G_{0,7}^{7,0}\left(\left.\frac{3.57k^4\ell^6\beta^7}{10^6}\right|
\begin{array}{c}
 -\frac{1}{6},0,\frac{1}{6},\frac{1}{4},\frac{1}{2},\frac{1}{2},\frac{3}{4} \\
\end{array}
\right)}$
& $\dfrac{2.5 \beta ^6 k^4 \ell^6 G_{0,7}^{7,0}\left(\left.\frac{3.57k^4\ell^6\beta^7}{10^6}\right|
\begin{array}{c}
 -\frac{7}{6},-\frac{5}{6},-\frac{3}{4},-\frac{1}{2},-\frac{1}{2},-\frac{1}{4},0 \\
\end{array}
\right)}{10^{5}G_{0,7}^{7,0}\left(\left.\frac{3.57k^4\ell^6\beta^7}{10^6}\right|
\begin{array}{c}
 -\frac{1}{6},0,\frac{1}{6},\frac{1}{4},\frac{1}{2},\frac{1}{2},\frac{3}{4} \\
\end{array}
\right)}$\\ & &\\ \hline & &\\
        $2$ & $\dfrac{\sqrt{k}\abs{\ell}}{2}$&$\dfrac{1}{\beta}+\sqrt{k}\abs{\ell}$ \\ & &\\ \hline & &\\
        
        $2.5$ & $\dfrac{5 G_{0,9}^{9,0}\left(\left.\frac{k^4\ell^{10}\beta^9}{10^9}\right|
\begin{array}{c}
 0,\frac{1}{10},\frac{1}{4},\frac{3}{10},\frac{1}{2},\frac{1}{2},\frac{7}{10},\frac{3}{4},\frac{9}{10} \\
\end{array}
\right)}{\beta  G_{0,9}^{9,0}\left(\left.\frac{k^4\ell^{10}\beta^9}{10^9}\right|
\begin{array}{c}
 -\frac{1}{10},0,\frac{1}{10},\frac{1}{4},\frac{3}{10},\frac{1}{2},\frac{1}{2},\frac{7}{10},\frac{3}{4} \\
\end{array}
\right)}$
&$\dfrac{9k^4\ell^{10}\beta^8 G_{0,9}^{9,0}\left(\left.\frac{k^4\ell^{10}\beta^9}{10^9}\right|
\begin{array}{c}
 -\frac{11}{10},-\frac{9}{10},-\frac{3}{4},-\frac{7}{10},-\frac{1}{2},-\frac{1}{2},-\frac{3}{10},-\frac{1}{4},0 \\
\end{array}
\right)}{10^{9} G_{0,9}^{9,0}\left(\left.\frac{k^4\ell^{10}\beta^9}{10^9}\right|
\begin{array}{c}
 -\frac{1}{10},0,\frac{1}{10},\frac{1}{4},\frac{3}{10},\frac{1}{2},\frac{1}{2},\frac{7}{10},\frac{3}{4} \\
\end{array}
\right)}$\\ & &\\ \hline& &\\

        $3$ & $\dfrac{3 G_{0,5}^{5,0}\left(\left.\frac{k^2 \ell^6 \beta ^5}{7776}\right|
\begin{array}{c}
 0,\frac{1}{6},\frac{1}{2},\frac{1}{2},\frac{5}{6} \\
\end{array}
\right)}{\beta  G_{0,5}^{5,0}\left(\left.\frac{k^2 \ell^6 \beta ^5}{7776}\right|
\begin{array}{c}
 -\frac{1}{6},0,\frac{1}{6},\frac{1}{2},\frac{1}{2} \\
\end{array}
\right)}$
&$\dfrac{5k^2\ell^6\beta^4 G_{0,5}^{5,0}\left(\left.\frac{k^2 \ell^6 \beta ^5}{7776}\right|
\begin{array}{c}
 -\frac{7}{6},-\frac{5}{6},-\frac{1}{2},-\frac{1}{2},0 \\
\end{array}
\right)}{7776  G_{0,5}^{5,0}\left(\left.\frac{k^2 \ell^6 \beta ^5}{7776}\right|
\begin{array}{c}
 -\frac{1}{6},0,\frac{1}{6},\frac{1}{2},\frac{1}{2} \\
\end{array}
\right)}$\\ & &\\ \hline
\end{tabular}
\end{center}
where $G_{p,q}^{\,m,n} \!\left(z\left|  \begin{matrix} a_1, \dots, a_p \\ b_1, \dots, b_q \end{matrix}\right.\right)$ is the Meijer G-function. Average potential energy, $\ev{V(r)}$ can be calculated using $\ev{V(r)}=\ev{E}-\ev{E_{\mathrm{rot}}}-T/2$.

\bibliographystyle{apsrev4-2}
%\bibliography{references.bib}
%